\documentclass[10pt,preprint]{sensys-proc}

\pdfoutput=1
\usepackage{times}
\usepackage{graphicx}
\usepackage{balance}
\usepackage{comment}
\usepackage{soul}
\usepackage{color}
\usepackage{amsfonts}
\usepackage{url}

\title{DSP.Ear: Leveraging Co-Processor Support for\\ Continuous Audio Sensing on Smartphones}

\author{
Petko Georgiev$^\S$, Nicholas D. Lane\textsuperscript{\textdagger}, Kiran K. Rachuri$^\S$\textsuperscript{\textdaggerdbl}, Cecilia Mascolo$^\S$\\
$^\S$University of Cambridge, \textsuperscript{\textdagger}Microsoft Research, \textsuperscript{\textdaggerdbl}Samsung Research America \\
}

\crdata{978-1-4503-3143-2}
\conferenceinfo{SenSys'14,} {November 3--6, 2014, Memphis, TN, USA.}
\CopyrightYear{2014}
\begin{document}

\maketitle

\begin{abstract}
The rapidly growing adoption of sensor-enabled smartphones has greatly fueled the proliferation of applications that use phone sensors to monitor user behavior. A central sensor among these is the microphone which enables, for instance, the detection of valence in speech, or the identification of speakers. Deploying multiple of these applications on a mobile device to continuously monitor the audio environment allows for the acquisition of a diverse range of sound-related contextual inferences. However, the cumulative processing burden critically impacts the phone battery.

To address this problem, we propose \textit{DSP.Ear} -- an integrated sensing system that takes advantage of the latest low-power DSP co-processor technology in commodity mobile devices to enable the {\em continuous and simultaneous} operation of multiple established algorithms that perform complex audio inferences. The system extracts emotions from voice, estimates the number of people in a room, identifies the speakers, and detects commonly found ambient sounds, while critically incurring little overhead to the device battery. This is achieved through a series of pipeline optimizations that allow the computation to remain largely on the DSP. 
Through detailed evaluation of our prototype implementation we show that, by exploiting a smartphone's co-processor, DSP.Ear achieves a $3$ to $7$ times increase in the battery lifetime compared to a solution that uses only the phone's main processor. In addition, DSP.Ear is $2$ to $3$ times more power efficient than a na\"{i}ve DSP solution without optimizations. We further analyze a large-scale dataset from $1320$ Android users to show that in about $80$-$90$\% of the daily usage instances DSP.Ear is able to sustain a full day of operation (even in the presence of other smartphone workloads) with a single battery charge.

\end{abstract}
\category{\noindent H.1.2}{User/Machine Systems}{Human information processing}
\terms{\noindent Design, Experimentation, Performance}

\section{Introduction}
\label{sec:intro}


In recent years, the popularity of mobile sensing applications that monitor user activities has soared, helped by the wide availability of smart and wearable devices. The applications enabled by the microphone have attracted increasing attention as audio data allows a wide variety of deep inferences regarding user behavior~\cite{Lee:2013:SEF:2462456.2465426,Lu:2012:SDS:2370216.2370270,Rachuri:2010:EMP:1864349.1864393,Xu:2013:CUS:2493432.2493435,pitchYin}. The tracking of one's activities requires a fairly continuous stream of information being recorded. Most smartphones already boast a quad-core or octa-core processor, therefore, computation is increasingly less of a constraint on these commonly used devices. However, the biggest limiting factor for smartphone applications that depend on sensor data is the inadequate battery capacity of phones. This problem is exacerbated in the case of the microphone sensor, due to its high data rates. 
Further, as more of these sensing applications become deployed on the device, the high computational requirements of simultaneously running complex classifiers add to the energy burden. As a result, duty cycling strategies are commonly adopted, but the side effect is a reduced coverage of captured sound events, since these can be missed while the sensor is not sampling.

Recently, a new architecture has been adopted by many smartphone manufacturers whereby an additional low-power processor, typically referred to as co-processor, is added alongside the more powerful primary processor.  The aim of this additional low-power processor is to sample and process sensor data. For example, the Apple iPhone~5S is equipped with an M7 motion co-processor~\cite{M7} that collects and processes data from the accelerometer and gyroscope even when the primary processor is in sleep mode. Similarly, the Motorola Moto X~\cite{MotoX} is equipped with a ``contextual computing processor'' for always-on microphone sensing but is limited to recognizing only spoken voice commands. A key challenge in leveraging such a co-processor is effectively managing its fairly limited computational resources to support a diverse range of inference types. 

In this work, we present a system that operates within the hardware constraints of low-power co-processors (Qualcomm Hexagon DSP \cite{hexagonDSP}) to continuously monitor the phone's microphone and to infer various contextual cues from the user's environment at a low energy cost. Most existing work (e.g.,~\cite{Lu:2011:SEE:2021975.2021992, LittleRock}) on using a low-power processor for sensing has employed custom hardware: we build our system on off-the-shelf phones and propose several novel optimizations to substantially extend battery life. Sensing with commodity co-processor support (e.g., \cite{Shen:2013:EPH:2525526.2525856,ra:ubicomp12}) is still exploratory. The computational constraints of the low-power units has limited the research to either fairly simple tasks such as accelerometer step counting or pipelines typically seen in isolation or in pairs. {\em We take these efforts one step further by studying how the co-processor in off-the-shelf smartphones can support multiple computationally intensive classification tasks on the captured audio data, suitable for extracting, for instance, human emotional states~\cite{Rachuri:2010:EMP:1864349.1864393} or stress~\cite{Lu:2012:SDS:2370216.2370270}}. In our design we are able to support the interleaved execution of five existing audio pipelines: ambient noise classification \cite{Lu:2009:SSS:1555816.1555834}, gender recognition \cite{pitchYin}, counting the number of speakers in the environment \cite{Xu:2013:CUS:2493432.2493435}, speaker identification \cite{Rachuri:2010:EMP:1864349.1864393}, and emotion recognition \cite{Rachuri:2010:EMP:1864349.1864393}. This is done \textit{primarily on the DSP itself} with only limited assistance from the CPU, maximizing the potential for energy efficiency.


The contributions of our work are:
\begin{itemize}
\item The design, and implementation, of a novel framework for smartphone microphone sensing (DSP.Ear) which allows continuous and simultaneous sensing of a variety of audio-related user behaviors and contexts. 
\item A detailed study of the trade-offs of CPU and co-processor support for sensing workloads. We design a series of techniques for optimizing interleaved sensing pipelines that can generalize to other future designs.
\end{itemize}

\vspace{+0.5em} \noindent We provide an extensive evaluation of the system and find that it is $3$ to $7$ times more power efficient than baselines running solely on the CPU. The optimizations we introduce prove critical for extending the battery lifetime as they allow DSP.Ear to run $2$ to $3$ times longer on the phone compared to a na\"{i}ve DSP deployment without optimizations. Finally, by analyzing a large 1320-user dataset of Android smartphone usage we discover that in $80$\% to $90$\% of the daily usage instances DSP.Ear is able to operate for the whole day with a single battery charge -- without impacting non-sensor smartphone usage and other phone operations. 
 
\section{Hardware Support for Continuous Sensing}
\label{sec:background}
Low-power co-processors for mobile devices are usually narrowly specialized to perform dedicated tasks in an energy friendly manner~\cite{Lu:2011:SEE:2021975.2021992}. Such processing units prove especially handy for continuous sensing where the sensor sampling power consumption costs~\cite{LittleRock} are several orders of magnitude lower than CPU-driven sensing designs that maintain a wake lock on the main CPU. However, co-processor use often introduces slower computation (due to the units being low-power and generally more limited in capabilities).

Two very popular co-processor units made available in high-end off-the-shelf smartphone models are the iPhone's M7 motion co-processor \cite{M7} and the Qualcomm's Hexagon QDSP6~\cite{Snapdragon} of the Snapdragon $800$ processor platform. Whereas the former is specialized solely to efficiently offload collection and processing of motion sensing through accelerometer, gyroscope, and compass, the latter excels at low-power multimedia processing. An important distinguishing feature between the two is that, unlike iPhone's M7, the Hexagon QDSP6 supports custom programmability. This is achieved through the publicly released C/assembly-based Hexagon SDK \cite{HexgaonSDK} which, to date, has been used for multimedia applications but has yet to be used for sensing. 


The realization of our targeted scenario of multiple concurrently running audio inference pipelines with emerging co-processor technology such as the Hexagon application DSP 
becomes particularly challenging because of the following design space limitations:

\vspace{+0.5em} \noindent \textbf{Memory Restrictions. }
The DSP runtime memory constraints restrict the amount of in-memory data reserved to sensing applications. Examples of such data are classification model parameters, accumulated inferences or features extracted for further processing. The lack of direct file system support for the DSP imposes the interaction with the CPU which will either save the inferences or perform additional processing on DSP-computed features. This easily turns into a major design bottleneck if the DSP memory is low, the data generated by multiple sensing applications grows fast, and the power-hungry CPU needs to be woken up from its low-power standby mode specifically to address the memory transfers.

\vspace{+0.5em} \noindent \textbf{Code Footprint. }
The DSP restricts the size of the shared object file deployed with application code. Deployment issues arise when machine learning models with a large number of parameters need to be initialized but these parameters cannot be read from the file system and instead are provided directly in code which exhausts program space. Sensing inferences performed on the DSP become restricted to the ones the models of which successfully fit under the code size limit.

\vspace{+0.5em} \noindent \textbf{Performance. }
The DSP has different performance characteristics. While ultra low-power is a defining feature of the DSP, many legacy algorithms which have not been specifically optimized for the co-processor hardware architecture will run slower. Computations that are performed in real time on the CPU may not be able to preserve this property when deployed on the DSP. The DSP generally supports floating point operations via a $32$-bit FP multiply-add (FMA) unit, but some of the basic operations such as division and square root are implemented in software which potentially introduces increased latency in some of the algorithms. 

\vspace{+0.5em} \noindent \textbf{Programability. }
The DSP supports only a subset of the symbols commonly found in C programming environments. This limits the range of already developed C libraries that can be ported to the DSP without modifications.

\section{System Overview}
\label{sec:overview}
In this section, we introduce the main architectural components of DSP.Ear as well as a high-level workflow of its operation. The system has been designed to perform the continuous sensing of a range of user behaviors and contexts, in near real-time, by leveraging the energy-efficient computation afforded by commodity DSPs found in a number of recent smartphones. To achieve this design goal we have addressed two fundamental challenges:

\vspace{+0.5em} \noindent \textbf{Supporting Multiple Complex Audio-based Inference Pipelines. } %
To recognize a broad set of behaviors and contexts we must support a variety of complex inference pipelines.  We have implemented a series of concurrently-operating representative audio-based pipelines based on previously published research: ambient noise classification \cite{Lu:2009:SSS:1555816.1555834}, gender recognition \cite{pitchYin}, speaker counting \cite{Xu:2013:CUS:2493432.2493435}, speaker identification \cite{Rachuri:2010:EMP:1864349.1864393}, and emotion recognition \cite{Rachuri:2010:EMP:1864349.1864393}. 

\vspace{+0.5em} \noindent \textbf{Operating within Commodity Co-Processor Mobile Architectural Limits. } %
To leverage the low-power co-processor capabilities of DSPs requires our design to cope with a number of architectural bottlenecks. The DSP can easily be overwhelmed by the high-sampling rates of the microphone and the bursts of computation needed to process audio data deep into pipelines, depending on context -- for example, when the user is in conversation requiring a number of inferences (such as emotion, speaker identification, gender estimation). 

\vspace{+0.5em} 
We overcome these challenges through our system architecture and implementation that interleaves the execution of five inference pipelines principally across a single standard DSP -- critically {\em our design enables each pipeline to be largely executed directly on the DSP and minimizes the frequency to offload computation to the primary CPU}.  
A key design feature is a series of pipeline execution optimizations that reduce computation through cross-pipeline connections in addition to leveraging common behavioral patterns. 

\begin{figure}
  \centering
  \includegraphics[width = \columnwidth]{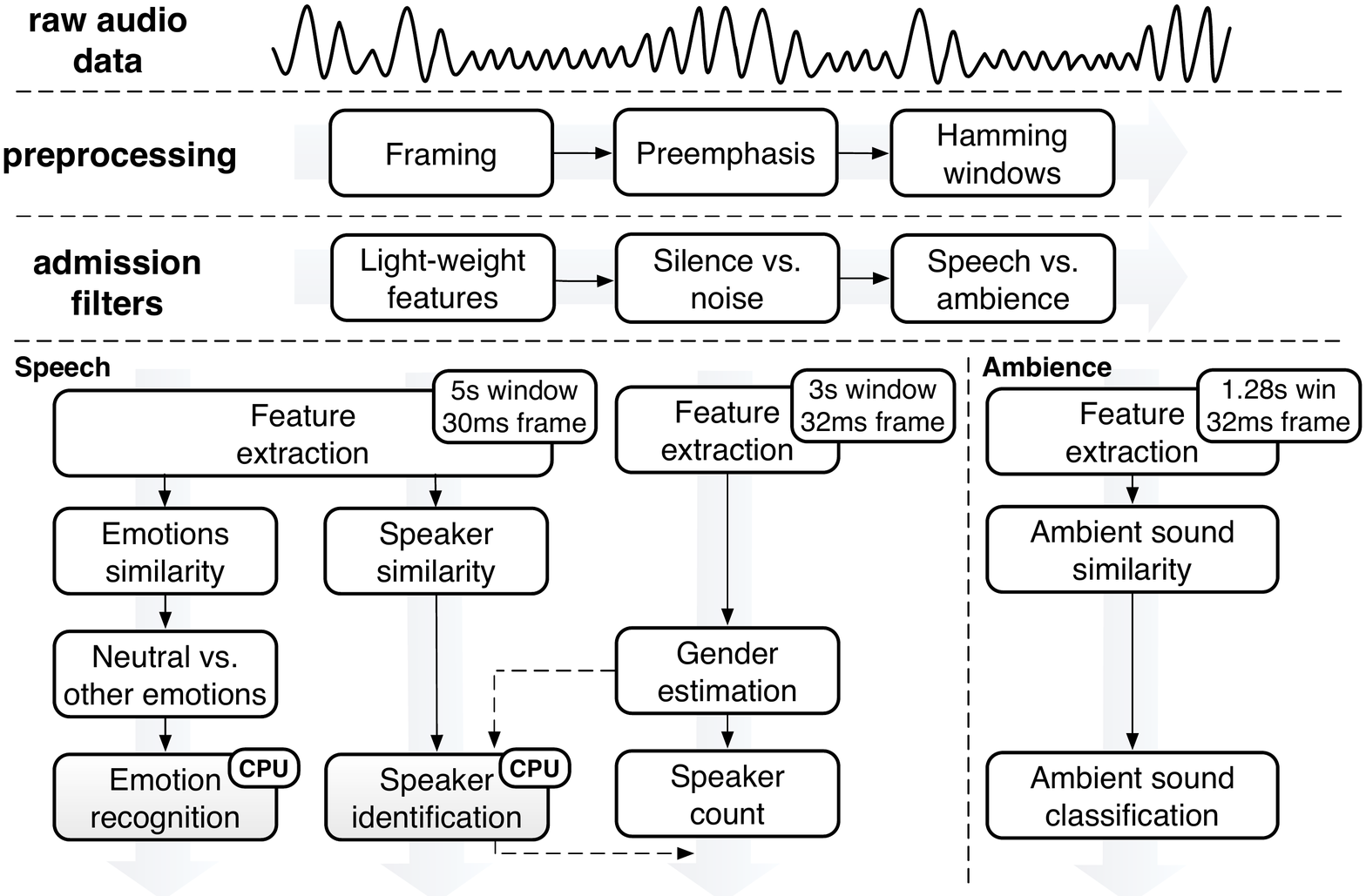}
  \caption{DSP.Ear Architecture.}
  \label{SystemOverview}
  \vspace{-0.5cm}
\end{figure}

Figure~\ref{SystemOverview} shows the overall system architecture. In DSP.Ear, the phone's microphone is continuously sampled (i.e., without any interruption) on the DSP, which applies a series of admission filters to the sampled audio. Light-weight features are then extracted to determine the presence of acoustic events. If the sampled audio passes an initial admission filter that filters out non-sound samples, then volume-insensitive features are extracted. Further, an additional filter splits the execution into two processing branches -- one for speech, and the other for ambient sounds -- depending on the output of a decision tree classifier.  
We now briefly describe each of these execution branches.  




\vspace{+0.5em} \noindent \textbf{Speech Processing Branch. } Under this pipeline branch we perform the following human voice-related inferences. 

\vspace{+0.5em} \noindent \textit{Gender Estimation. } A binary classification is performed to identify the gender of the speaker. The output of this classifier also assists with subsequent pipeline stages that estimate the number of nearby people (Speaker Count) and recognizing which person is speaking (Speaker Identification).  

\vspace{+0.5em} \noindent \textit{Speaker Count. } To find the number of speakers in a conversation, we implement an adapted version of the Crowd++ unsupervised algorithm \cite{Xu:2013:CUS:2493432.2493435}. We continuously extract features from 3-second long utterances and perform on-stream merging of subsequent segments into the same cluster depending on whether the feature vectors are close enough. Once we detect that the conversation has stopped after a minute of no talking, we schedule the final stage of the speaker counting algorithm.

\vspace{+0.5em} \noindent \textit{Emotion Recognition. } We divide the audio stream into 5-second long samples and on the DSP extract a different feature set required by the emotion and speaker identification stages. A light-weight similarity detector computes summary statistics (mean and variance) over the extracted acoustic features and compares the summary vectors of the previous and current audio segment. If the features are similar enough, the emotion labels are propagated from the previous inference.

The emotion classification stage consists of several steps and is mainly performed on the CPU. The first step is applying an admission filter that accounts for whether the emotion is neutral or not. It operates faster than running the full pipeline of all emotions and often saves computation given that neutral emotions dominate in everyday settings \cite{Rachuri:2010:EMP:1864349.1864393}. If the emotion is not neutral, the finer category classification continues with determining the narrow non-neutral emotion such as happiness, sadness or fear. 

\vspace{+0.5em} \noindent \textit{Speaker Identification. } The speaker identification algorithm uses the same set of features required for the emotion recognition task and again makes inferences based on audio recordings of the same length (5 seconds). An identical similarity detector calibrated with a different similarity threshold eliminates redundant classifications when the cosine angle between summary feature vectors of subsequent audio recordings is sufficiently low. 
In the classification stage when the likelihood of offline trained speaker models is derived for the audio recordings, the already estimated genders are cross-correlated with the models to reduce the number of computed likelihoods to the set of gender-speaker matches.

\vspace{+0.5em} \noindent \textbf{Ambient Sound Processing Branch}. This pipeline branch deals with the detection of everyday sounds. We train in an offline manner several classifier models that represent examples of commonly encountered sounds in natural environments (music, water, traffic, other). The classifiers require the computation of an additional set of summary features over a commonly adopted $1.28$-second window \cite{Lu:2012:SDS:2370216.2370270, Xu:2013:CUS:2493432.2493435}. At runtime, an ambient sound similarity detector intercepts the classification process and firstly compares the feature vectors of subsequent windows for similarity. If the acoustic fingerprints are sufficiently close to each other, the same type of sound is inferred which bypasses the expensive classification process. In the alternative case, the final stage of the ambient processing pipeline consists of finding the sound model that with the highest probability matches the input sequence of acoustic observations. This classification step becomes more expensive with the increase in the number of sounds against which the audio input is being matched.
%


\section{Audio Inference Pipelines}
\label{sec:algorithms}
In this section, we detail five microphone-based behavioral inference pipelines currently included in our system.  Four of these pipelines are under the speech processing branch while one is in the ambient processing branch.
Our design is extensible and additional audio pipelines (e.g., SocioPhone~\cite{Lee:2013:SEF:2462456.2465426}, StressSense~\cite{Lu:2012:SDS:2370216.2370270}) can easily be added as required to capture additional dimensions of the user behavior and context.  
Each implemented pipeline in our system is based on previously published research.  
Table~\ref{table:features} provides an overview of the features and inference algorithms used.
We now describe each pipeline in turn.

\begin{table}[t]
\centering
\small
\begin{tabular}{|p{5cm}|l|}
\hline
\textbf{Feature Set} & \textbf{Pipelines and Filters} \\
\hline
\hline

RMS, Spectral Entropy \cite{Lu:2009:SSS:1555816.1555834} & Silence Filter \\
\hline
pitch \cite{pitchYin} & Crowd++, Gender \\
\hline
MFCC \cite{mfcc} & Crowd++, Ambient \\
\hline
PLP ($16$ static and $16$ delta) \cite{plp} & Emotions, Speaker Id  \\
\hline
Low Energy Frame Rate \cite{lefr}, Zero Crossing Rate \cite{zcr}, Spectral Flux \cite{lefr}, Spectral Rolloff \cite{feats}, Spectral Centroid \cite{feats}, Bandwidth \cite{feats}, Relative Spectral Entropy \cite{Lu:2009:SSS:1555816.1555834}, Normalized Weighted Phase Deviation \cite{Dixon06} & Speech Filter, Ambient \\
\hline
\end{tabular}
\vspace{+0.1cm}
\caption{Acoustic features.}
\label{table:features}
\vspace{-0.1cm}
\end{table}

\vspace{+0.5em} \noindent \textbf{Emotion Recognition}. Emotions are an integral part of a user's everyday life and to detect them we analyze human voice by employing the algorithm introduced by EmotionSense \cite{Rachuri:2010:EMP:1864349.1864393}. Gaussian Mixture Model classifiers with diagonal covariance matrices are trained using emotional speech from the Emotional Prosody Speech and Transcripts library \cite{emotionsLibrary}. Each of 14 narrow emotions (Table~\ref{table:emotions}) is represented by an emotion-specific GMM classifier built by performing Maximum a Posteriori (MAP) adaptation of a 128-component background GMM representative of all emotional speech. 

\begin{table}[t]
\centering
\small
\begin{tabular}{|l|p{5.7cm}|}
\hline
\textbf{Broad Emotion} & \textbf{Narrow Emotions} \\
\hline
\hline
Anger & Disgust, Dominant, Hot Anger  \\
\hline
Fear & Panic \\
\hline
Happiness & Elation, Interest, Happiness \\
\hline
Neutral & Boredom, Neutral Distant, Neutral Conversation, Neutral Normal, Neutral Tete, Passive \\
\hline
Sadness & Sadness \\
\hline
\end{tabular}
\vspace{+0.2cm}
\caption{Emotion categories (adopted from \cite{Rachuri:2010:EMP:1864349.1864393}).} 
\label{table:emotions}
\vspace{-0.5cm}
\end{table}

The GMM evaluates the probability of certain observations being attributed to the model and in our case these observations are Perceptual Linear Prediction (PLP) coefficients \cite{plp} extracted from frames over $5$ seconds of recorded audio. The audio signal is segmented into $30$ms frames with a $20$-ms overlap and 32 PLP coefficients are computed from each frame. At runtime, the likelihood of the recorded audio sequence is calculated given each emotion class model and the emotion corresponding to the highest likelihood is assigned. As described by the authors \cite{Rachuri:2010:EMP:1864349.1864393}, the narrow emotions are grouped together into 5 broad categories that capture sadness, happiness, fear, anger and neutral speech. The final result of the recognition process is thus the broad category to which the classified narrow emotion belongs.

\vspace{+0.5em} \noindent \textbf{Speaker Identification}. The speaker recognition algorithm is again based on~\cite{Rachuri:2010:EMP:1864349.1864393} and it reuses the algorithmic elements introduced in the emotion classification. A 128-component background GMM representative of all available speakers is built and MAP adaptation is performed on PLP features from speaker specific utterances to obtain the speaker-dependent models. At runtime, the speaker-specific model that produces the highest likelihood given the audio sequence is identified as the speaker of the audio sample. 


\vspace{+0.5em} \noindent \textbf{Gender Estimation}. One of our use cases is determining the gender of the current speaker. Previous studies \cite{gender} have demonstrated that the most distinctive trait between male and female voices is their fundamental frequency, also known as pitch. Similarly to Xu et al.~\cite{Xu:2013:CUS:2493432.2493435}, we adopt Yin's algorithm \cite{pitchYin} to determine the pitch from a $32$-ms frame. We compute a series of pitch values from the $50\%$ overlapping frames in a 3-second long window and use the mean as a summary statistic over the whole utterance. Gender inferences are made based on the findings of Baken \cite{gender} that the average pitch for men typically falls between $100$ and $146$Hz, whereas for women it is usually between $188$ and $221$Hz. The algorithm we use infers a male voice for the whole utterance if the pitch is below $160$Hz, reports a female voice if the pitch is above $190$Hz and is uncertain in the cases where the value falls between these thresholds.

\vspace{+0.5em} \noindent \textbf{Speaker Count}. We count the number of speakers in a conversation by using the Crowd++ unsupervised counting algorithm~\cite{Xu:2013:CUS:2493432.2493435}. It consists of 2 phases: feature extraction and speaker counting. In the first phase, speech is segmented into 3-second long windows and 20-dimensional MFCC features are extracted from the $32$-ms frames with $50\%$ overlap. The speaker counting phase is triggered infrequently and when the conversation is over. A forward pass stage, linear in the number of segments, merges neighboring segments into clusters represented by the mean values of the MFCC features. Two segments are merged together if the cosine angle between the representative feature vectors of the segments falls below an experimentally identified threshold ($15^{\circ}$). Thus clusters correspond to audio sequences during which the same speaker is talking. Once this is done, the final stage of the algorithm compares clusters against each other and merges them based on their cosine similarity and the inferred gender of the speaker represented by the cluster. The total number of identified clusters in this final stage is the inferred number of speakers in the conversation.

\vspace{+0.5em} \noindent \textbf{Ambient Sound Classification}. For the detection of various ambient sounds we again adopt Gaussian Mixture Models which have proven effective in discriminating between human activity sounds in natural environments \cite{ambient}. We divide the audio signal into $32$-ms non-overlapping frames and extract features (Table~\ref{table:features}) from each frame in a window of $40$ frames. The audio sample length of $1.28$ seconds is small enough to account for short-length sounds, while at the same time being wide enough to capture distinctive acoustic characteristics. The series of frame features in the window constitute the acoustic observations which are then evaluated against the GMM models. We pick 3 commonly encountered types of sounds (music, traffic, and water) as some examples and build a total of 4 GMMs representative of the 3 sound categories and other noises. We use $66$ as the number of mixture components after evaluating the Bayesian Information Criterion (BIC)~\cite{bic} on several validation models. The GMMs are trained with a standard expectation maximization (EM) algorithm.

\section{Concurrent Inference Pipeline Support}
\label{sec:optimizations}

In what follows, we describe four general categories of audio inference pipeline optimizations designed to enable each pipeline to operate concurrently within the (largely) hardware limitations of our target prototype platform. These optimization categories include: \textit{admission filters} enabling the early elimination of unnecessary pipeline stages; \textit{behavioral locality detection} for reducing the frequency of full inference computation; \textit{selective CPU offloading} for delegating tasks the DSP is unable to process; and \textit{cross-pipeline optimizations}. 

\subsection{Admission Filters}
We adopt three distinct admission filters based on the combination of implemented audio inference pipelines.

\vspace{+0.5em} \noindent \textbf{Silence filtering. } %
A large proportion of the time users are situated in silent environments where the ambient noise is low. In such cases performing audio processing is a waste of phone resources. Similarly to \cite{Lu:2009:SSS:1555816.1555834} we divide the audio stream into a series of frames and compute the Root Mean Square (RMS) and spectral entropy which we test against experimentally determined thresholds to decide whether the frame is silent or not. We use a shorter frame of $32$ms which allows us to increase the precision with which we detect the onset of sounds. In a manner similar to \cite{Lu:2011:SEE:2021975.2021992} this stage of the processing can be delegated to the low-power co-processor and thus eliminate the need for adopting duty cycling schemes which may miss sound events. Once the frame is determined to contain some acoustic event, all subsequent frames in a time window are admitted for further processing. If a time window occurs such that all frames inside are flagged as silent, frame admission ceases.

\vspace{+0.5em} \noindent \textbf{Neutral emotion biasing. } 
As reported by \cite{Rachuri:2010:EMP:1864349.1864393} between $60$\% and $90\%$ of the time the emotions encountered in human speech are neutral. This finding implies that being able to quickly flag an utterance as being neutral or not could provide significant savings in processing power. In most cases the binary decision would result in the emotion being classified as neutral which can bypass the time-consuming comparisons against all emotion class models. We therefore build 2 GMMs, one representative of neutral emotions and a filler one capturing the rest. When the emotion recognition part ensues we first perform a classification against these two models and finer emotion recognition proceeds only if the detected valence is not neutral.

\vspace{+0.5em} \noindent \textbf{Speech detection. } %
The majority of the application scenarios we consider require the analysis of human voice which is why ensuring that such processing occurs only when speech is encountered is mandatory. The coarse category classification of the audio samples into speech and ambient noise is a frequently performed step that occurs whenever the environment is not silent. It needs to be fast and efficient while at the same time retaining high accuracy to reduce the false positive rate and avoid unnecessarily triggering the expensive speech processing pipelines. We adopt a strategy that has been established in previous works \cite{Lu:2009:SSS:1555816.1555834, Lu:2011:SEE:2021975.2021992} where we use non-overlapping frames in a window from which we extract features and perform binary classification via a J$48$ decision tree on the whole audio window. We use $32$ms as the frame length and a window of $40$ frames which amounts to $1.28$ seconds of audio sampling before deciding whether the sound sample contains speech or ambient noise. The low energy frame rate \cite{lefr} as well as the mean and variance over a window of the features shown in Table~\ref{table:features} are used. 

\subsection{Leveraging Behavioral Locality in Audio}
Many human activities and behaviors, such as taking a shower or being in a specific emotional state, last much longer than a few seconds. However, conventionally our inference pipelines perform the entire inference process each time a new audio data segment arrives from the microphone -- even though often the prior user activity or context will be still on-going. The largest computational waste is the incremental Bayesian process of testing each GMM model used to represent possible behavior or sound categories to find the most likely model given the data segment. Instead, similarly to Lu et al. \cite{jigsaw2010}, we adopt a scheme that computes the similarity between the current and previous feature vectors and triggers the full classification step only if the new acoustic signature differs significantly from the previous one. 
For Ambient Sound Classification we use summary values (mean and variance) of the features to build vectors that will be compared from subsequent windows. We compute the cosine angle between these vectors and if it remains within a threshold $\delta$ we do not run the full classification pipeline and instead propagate the prior category.
For Emotion and Speaker Recognition we compute the mean and variance of the PLP coefficients in an audio sample and compare subsequent segments for similarity. If these derived vectors are similar enough we reuse the label of the previous recognized emotion or speaker. Note that this strategy is reasonable since the emotional state is unlikely to oscillate violently between subsequent speech utterances.




\subsection{Selective CPU Offloading}
\label{sec:offloading}
There are several important parameters that drive the design of an integrated system for audio processing running on the DSP. On the one hand, there is a limit on the amount of memory allocated for the compiled code that will be deployed on the DSP. Since the co-processor does not have a file system of its own, the model parameters for the GMMs cannot be loaded from a file but need to be initialized directly through code. This restricts the amount of sounds, emotions and speakers that can be recognized on the co-processor. Therefore, even in the cases when we could afford to opportunistically perform heavy and time-consuming classifications on the DSP, there will be a limit on the type of inferences we could make. This constraint drives the need for selectively offloading computation on the main CPU where it could perform the tasks the DSP is unable to handle.

To accomplish this offloading the DSP needs to interact with the CPU to transfer feature buffers for further processing. Often the CPU will be in sleep mode at the point of DSP interaction initiation which means that the DSP needs to wake up the CPU. This is currently supported on the Snapdragon 800 platform through a mechanism known as FastRPC that is based on remote procedure calls. The CPU invokes a remote method on the DSP where the execution can be blocked on a variable until the DSP updates its status. Due to the thread migration mechanism implemented by Qualcomm, during the method invocation the CPU can go to sleep if the DSP execution takes too long. On returning from the remote method, the CPU is woken up automatically. The buffer transfers in this case consist of simply passing an argument to the remotely invoked method which is why the transfer itself is relatively inexpensive compared to the wake-up operation.

The rate at which the CPU is woken up is critical to the energy profile of the full system as waking the CPU is accompanied with a power consumption overhead originating from two sources. First, the wake-up itself consumes power that is an order of magnitude higher than the standby power. Second, once the buffer transferring is over, the CPU remains idle for some time before going back to sleep/standby mode. Therefore, reducing the wake-up rate proves to be of utmost importance for maintaining a low-energy profile. 

The time between two subsequent buffer transfers from the DSP to the CPU defines the CPU wake-up rate. This time is dependent on several factors, the most pronounced of which are the runtime memory limitations of the DSP as well as the frequency of encountered sound events. As the DSP can only accumulate a certain amount of data before waking up the CPU to offload computation, the offloading must work in synergy with the similarity detectors to reduce the number of transferred acoustic features and increase the number of inferences made by the DSP. The speech features computed over 5 seconds of audio recording consume 2 orders of magnitude more memory than the ambient features which means that the rate at which the CPU is woken up is largely determined by the proportion of the time speech is detected. We can formalize the time $\Delta t$ in seconds between two DSP-CPU interactions by the following equation:

\begin{displaymath}
\Delta t = \gamma + \frac{(M_L - M_M)}{M_P} \times (1 + \min(S_E, S_S)) \times \tau
\end{displaymath}

\noindent where, 
\begin{itemize}
\item
$\gamma$ is the time spread in seconds during which silence and ambient sounds interleave the speech;
\item
$M_L$ is the DSP memory limit;
\item
$M_M$ is the memory consumed by the audio module parameter initialization, the pipeline inferences plus the accumulated ambient features;
\item
$M_P$ is the size of the speech (PLP) features;
\item
$S_E$ and $S_S$ are the fractions of similar emotions and speakers identified by the similarity detectors;
\item
$\tau$ is the sampling window in seconds over which speech features are extracted (currently equal to $5$ seconds).
\end{itemize}

\vspace{+0.5em} \noindent 
Note that we need to take the minimum of $S_E$ and $S_S$ as the speech features are shared by the two classification algorithms. To give a perspective on this time $\Delta t$, if we assume that $\gamma$ is $0$ (no silence/ambience in between the speech), no classifications are saved and the memory limit is $8$MB we easily run out of co-processor space in around $9$ minutes. On the other hand, if we encounter only ambience and no speech we can accumulate ambient features for well over $9$ hours in the absence of silence.
To summarize, the frequency with which a user is involved in conversations is a prime determiner of how often we resort to selective CPU offloading. Maximizing the time between subsequent DSP-CPU interactions is crucial for maintaining a low-power system profile.

\subsection{Cross-Pipeline Optimizations}
Our design leverages the following inter-pipeline links to allow contextual hints extracted by early simpler pipeline components to benefit later more complex inference phases.

\vspace{+0.5em} \noindent \textbf{Gender Filtering for Speaker Identification. } We make use of the gender estimation pipeline to reduce the number of GMMs against which we perform speaker identification. If the gender of the speaker whose GMM is being evaluated does not match the inferred gender from the audio sequence we do not compute the probability for that model.

\vspace{+0.5em} \noindent \textbf{Speaker Count Prior from Speaker Identification. } We boost the accuracy of estimating the number of nearby people with a prior based on the number of unique speakers found by Speaker Identification. From Speaker Identification the number of speakers is estimated for those individuals our system possesses a speaker model. This is used to set a prior on the likelihood of each Speaker Count category of nearby crowd size. If Speaker Identification recognizes an unknown voice (i.e., a voice too dissimilar from all available speaker models) then category priors are adjusted to reflect the potential for additional nearby people.

\vspace{+0.5em} \noindent \textbf{Speech Detection activated Speaker Identification and Speaker Count. } Only if speech is detected by an admission filter are later complex inferences of Speaker Count and Speaker Identification made. Otherwise these pipelines of audio analysis are short circuited and never performed.


\section{Prototype Implementation}
\label{sec:implementation}
\begin{figure}[t] 
		\centering
        \begin{minipage}[b]{0.3\textwidth}
                \includegraphics[width=\textwidth]{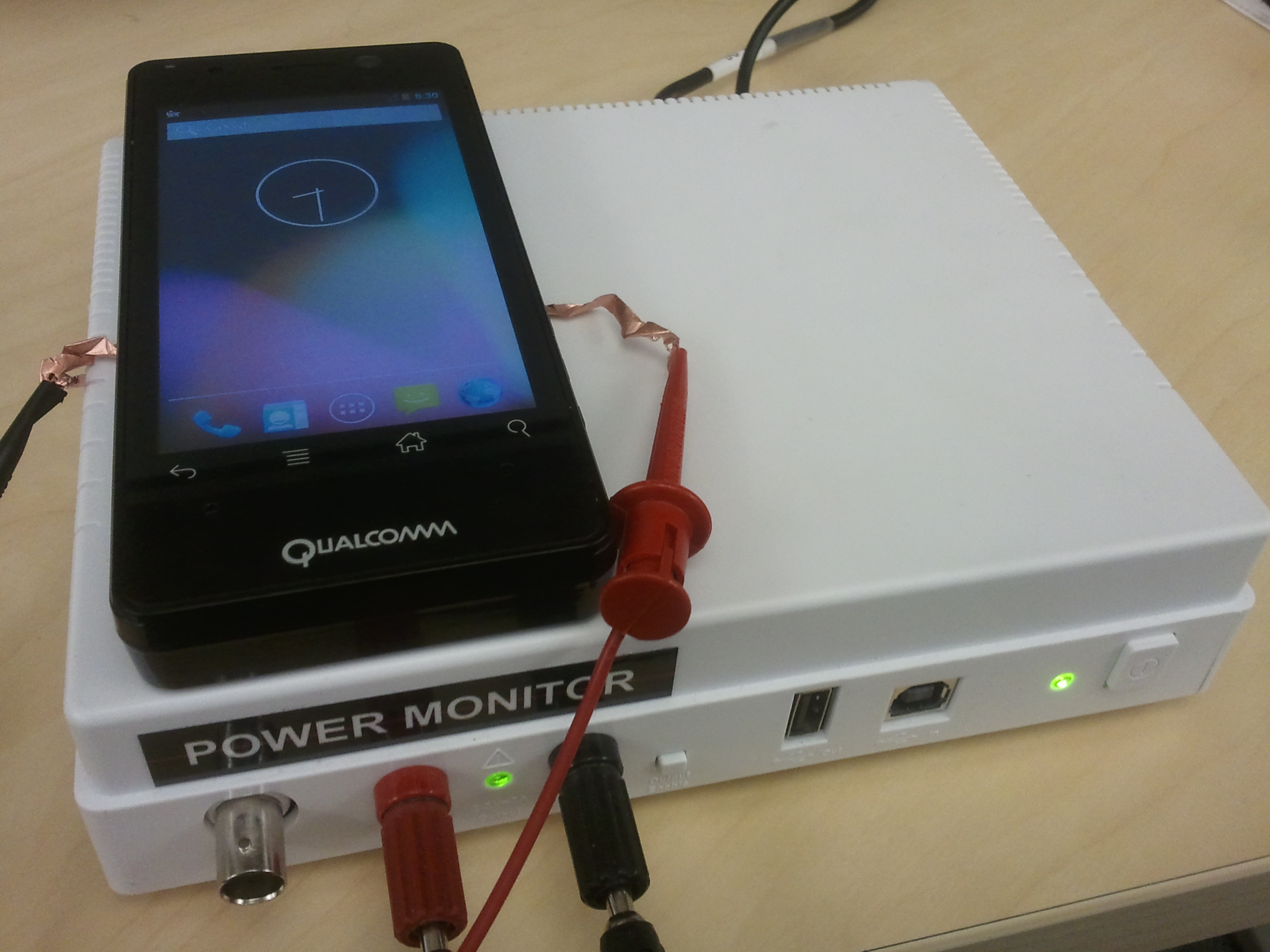}
        \end{minipage}%
        \caption{Snapdragon 800 Mobile Development Platform (MDP) \cite{mdp} used for the system development.} 
        \label{fig:mdp}
        \vspace{-0.5cm}
\end{figure}

The system prototype is implemented on a Snapdragon 800 Mobile Development Platform for Smartphones (MDP/S) with an Android Jelly Bean OS~\cite{mdp} (Figure~\ref{fig:mdp}). Access to the low-level co-processor APIs is granted through the C-based Hexagon SDK of which we use version $1.0.0$. The audio processing algorithms are implemented in C through the Elite firmware framework which is part of the SDK and is designed for the development of audio modules. We duplicate the C functionality of the audio processing for the Android OS where we utilize the Native Development Kit (NDK) to interface with the Java code. This is needed so that we can compare the system performance and efficiency against a CPU-only implementation. The audio sampling of the microphone on the CPU is performed in Java.

The DSP programmability is open to selected development devices such as the MDP/S but not to commodity smartphones featuring the same Snapdragon 800 processor. Currently the version of the C programming language supported on the DSP includes only a subset of the standard libraries commonly found in recent C compiler implementations. This drives the need for porting and modifying audio processing code from other libraries specifically for the DSP. We adapt common algorithms such as Fast Fourier Transforms, feature implementations and GMM classification from the HTK Speech Recognition Toolkit (HTK)~\cite{htk}. The training of the GMMs for the emotion and speaker recognition models is performed in an offline manner through the HTK toolkit, while the ambient mixture models are trained through the scikit-learn Python library~\cite{scikit}. The microphone sampling rate used for all applications is $8$kHz.

The audio module is deployed via a compiled shared object file with the system code. For our version of the SDK the maximum allowed size of this shared object file on the DSP is $2$MB. This introduces limits to the number of mixture models that could be kept on the DSP at runtime. Since the DSP does not have a file system of its own, the model parameters cannot be loaded from a file but need to be initialized directly through code. The compiled code for one emotion or ambient model occupies approximately $260$KB or $87$KB of the shared object file respectively. This leads to a maximum of $5$ emotion or $16$ ambient mixture models that could be used at runtime by the DSP given that the code for the functioning of the integrated system also requires space. Our prototype keeps loaded on the DSP $2$ GMMs for the "neutral vs. all" emotion admission filter and $4$ ambient mixture models as some examples of commonly found sounds in everyday life (music, traffic, water and a sink model capturing other sounds).

To enable real-time analysis of the audio stream we need to bring together the execution for the processing of the same type of sound. When human voice is detected we would like to make gender inferences, count the speakers and extract features needed for emotion and speaker recognition at the same time. Not doing this online as the microphone is being sampled would result in spending additional memory for keeping raw audio buffers for further processing at a later stage. With the relatively limited amount of memory available for the DSP (currently $8$MB) this case is undesirable. Therefore, we take advantage of the hardware threads on the DSP to enable multi-threaded execution. 

There are a total of three hardware threads on the DSP of which we effectively use two. The hardware threads unlike typical software ones are architected to look like a multi-core with communication through shared memory. 
Currently the speaker count algorithm is executed on-the-fly in the main processing thread where the audio buffers become available. The PLP feature extraction required for emotion and speaker identification, as well as the neutral emotions admission filter, are performed together in a separate thread as soon as $5$ seconds of audio recording are accumulated. Whereas enabling multi-threading naturally consumes more power in mW than running each pipeline individually, the latency is reduced so that the overall energy consumption in mJ remains roughly equal to the case of running the pipelines sequentially. This observation confirms a near perfect power scaling for the multi-threaded support of the DSP and it is a crucial feature for energy efficiency advertised by Qualcomm \cite{hexagonDSP}.


\section{Evaluation}
\label{sec:benchmarks}
In this section we provide an extensive evaluation of the proposed system and its various components. The main findings can be summarized to the following:
\begin{itemize}
\item
The only runtime bottlenecks on the DSP are the classification stages of the emotion and speaker recognition.
\item
Our design is between $3$ and $7$ times more power efficient than CPU-only baselines.
\item
The optimizations are critical for the extended battery lifetime of the system as they allow the DSP+CPU solution to operate $2$ to $3$ times longer than otherwise possible.
\item
Under common smartphone workloads the system is able to run together with other applications for a full day without recharging the battery in about $80$-$90$\% of the daily usage instances.
\end{itemize}

\vspace{+0.5em} \noindent In Section ~\ref{sec:micro_benchmarks} the evaluation highlights the accuracy, runtime and power profiles of the  pipelines in isolation. Section ~\ref{sec:optimization_benchmarks} details a study on the parameters and importance of the introduced optimizations. Last, Section ~\ref{sec:full_eval} gives an evaluation of the full system energy consumption compared against three baseline models.

\subsection{Inference Pipeline Micro-Benchmarks}
\label{sec:micro_benchmarks}
Here we focus on testing the individual audio pipelines with regard to the accuracy, runtime and power consumption characteristics. All measurements performed on the DSP are reported for a default clock frequency and a Release version of the deployed code. We show that among the pipelines the emotion and speaker recognition are a processing bottleneck for the DSP, whereas the feature extraction stages, ambient classification and speaker counting can be run efficiently and in near real time on the DSP. Finally, most application algorithms are an order of magnitude more power efficient when executed on the DSP.

\subsubsection{Datasets}
We evaluate our implemented pipelines using the following datasets.

\vspace{+0.5em} \noindent \textbf{Emotions}: Similarly to EmotionSense \cite{Rachuri:2010:EMP:1864349.1864393} we use training and testing data from the Emotional Prosody Speech and Transcripts library \cite{emotionsLibrary}. The dataset consists of voiced recordings from professional actors delivering a set of 14 narrow emotions grouped into 5 broad categories (happiness, sadness, fear, anger and neutral). 

\vspace{+0.5em} \noindent \textbf{Speaker identification}: We use $10$-minute speech samples recorded by a total of $22$ speakers working in our research department at the time of the dataset collection.

\vspace{+0.5em} \noindent \textbf{Conversations and gender}: We extract $24$ minutes worth of conversational speech in various contexts from online radio programs. $12$ male and $12$ female voices are captured in natural turn-taking situations with occasional pauses.

\vspace{+0.5em} \noindent \textbf{Ambient sounds}: The dataset consists of $40$ minutes of various sounds equally split into the $4$ categories music, traffic, water and other. The music audio clips are a subset of the GTZAN genre collection \cite{Li02musicalgenre}; the traffic samples were downloaded from an online provider of free sound effects \cite{freesfx}; the water samples were obtained from the British Library of Sounds~\cite{britishLibrary}; the rest of the sounds were crawled from a subset of the SFX dataset \cite{Chechik:2008:LCA:1460096.1460115}. 

\vspace{+0.5em} \noindent \textbf{Speech vs. ambient noise}. We assemble a dataset that consists of $12$ minutes of conversations from online radio programs and $12$ minutes of various ambient sounds including street noise, traffic, water, weather effects, animal pet sounds, machinery, typing, and more. The sources of the ambient noise are as described in the previous paragraph.

\subsubsection{Accuracy}
As shown below, we discover the accuracy of each implemented pipeline is in line with already published results. We report the performance of the algorithms for correctness and with datasets recorded in relatively clean environments. Nevertheless, detailed analysis of the algorithmic accuracy under more challenging conditions can be found in the original papers \cite{Rachuri:2010:EMP:1864349.1864393,Xu:2013:CUS:2493432.2493435,jigsaw2010}. 

\vspace{+0.5em} 
\noindent \textbf{Speech detection.} In Table~\ref{table:speech} we show the confusion matrix of the decision tree classifier that distinguishes between speech and other types of sounds. It achieves an overall accuracy of $91.32\%$.

\begin{table}[t]
\centering
\small
\begin{tabular}{rrr}
& \textbf{Speech} & \textbf{Ambient Noise} \\
\hline
\hline
\textbf{Speech} & 93.02\% & 6.98\% \\
\textbf{Ambient Noise} & 10.34\% & 89.66\% \\
\hline
\end{tabular}
\vspace{+0.2cm}
\caption{Confusion matrix for speech/ambient noise classification. Results obtained via 5-fold cross validation.}
\label{table:speech}
\vspace{-0.5cm}
\end{table}

\vspace{+0.5em}  \noindent \textbf{Emotion recognition and speaker identification}. 
We confirm through our implementation an overall accuracy of $70.88\%$ for discriminating between the five broad emotion categories when acoustic features are computed over $5$ seconds of voiced speech \cite{Rachuri:2010:EMP:1864349.1864393}. Similarly to the reported results for the original implementation of the speaker identification, we observe an increase in the accuracy which on our dataset of $22$ speakers reaches $95\%$. 


\vspace{+0.5em} \noindent \textbf{Gender classification}. 
Table~\ref{table:gender} shows the confusion matrix for our implementation of the gender classification algorithm described by Xu et al. \cite{Xu:2013:CUS:2493432.2493435}. It is reasonably accurate with the male and female voices correctly classified as such $95.05\%$ and $90.63\%$ of the time respectively. The largest error in the classification is observed when the algorithm is uncertain about the gender of the detected voice, while errors caused by misclassification barely exceed $1\%$.

\begin{table}[t]
\centering
\small
\begin{tabular}{rrrr}
& \textbf{Male} & \textbf{Female} & \textbf{Uncertain} \\
\hline
\hline
\textbf{Male} & 95.05\% & 0.00\% & 4.95\% \\
\textbf{Female} & 1.04\% & 90.63\% & 8.33\% \\
\hline
\end{tabular}
\vspace{+0.2cm}
\caption{Confusion matrix for gender estimation based on pitch.}
\label{table:gender}
\vspace{-0.5cm}
\end{table}

\vspace{+0.5em}  \noindent \textbf{Speaker count}. We extract parts of the conversations in the described dataset and vary the number of speakers from 1 to 10 in a manner similar to Crowd++ \cite{Xu:2013:CUS:2493432.2493435}. The metric used by the authors to compute the accuracy of the speaker counting algorithm is Average Error Count Distance (AECD). It measures the error rate as the absolute difference between the actual and reported number of speakers. On the dataset we have gathered the Crowd++ algorithm achieves an AECD of $1.1$ which is consistent with the reported results for private indoor environments.

\vspace{+0.5em} \noindent \textbf{Ambient sound classification}. In this task we disambiguate between 3 distinct classes of sounds with fairly unique characteristics which together with the clean dataset might account for the high discriminating power reported in Table~\ref{table:ambient}. The main source of confusion comes from the other types of sounds which might be acoustically close to either music, traffic or water. For instance, some of the sounds in the \emph{Other} dataset are ring tones or chirping of birds which might be interpreted as music.

\begin{table}[t]
\centering
\small
\begin{tabular}{rrrrr}
& \textbf{Music} & \textbf{Traffic} & \textbf{Water} & \textbf{Other} \\
\hline
\hline
\textbf{Music} & 97.22\% & 0.22\% & 0.43\% & 2.13\% \\
\textbf{Traffic} & 0.27\% & 97.63\% & 1.05\% & 1.05\% \\
\textbf{Water} & 1.33\% & 1.55\% & 94.47\% & 2.65\% \\
\textbf{Other} & 8.42\% & 4.08\% & 5.36\% & 82.14\% \\
\hline
\end{tabular}
\vspace{+0.2cm}
\caption{Confusion matrix for ambient sound detection. Results are obtained through 5-fold cross validation.}
\label{table:ambient}
\vspace{-0.5cm}
\end{table}

\begin{table}[t]
\centering
\small
\begin{tabular}{rrr}
& \textbf{CPU} & \textbf{DSP} \\
\hline
\hline
\textbf{Silence} & 7.82 ms & 45.80 ms \\ 
\textbf{Speech} & 11.20 ms & 66.42 ms \\
\hline
\end{tabular}
\vspace{+0.2cm}
\caption{Normalized runtime for processing 1 second of audio data for the silence and speech admission filters.}
\label{table:runtime_filters}
\vspace{-0.4cm}
\end{table}

\subsubsection{Latency}

\begin{figure}[t]  
        \begin{minipage}[b]{0.23\textwidth}
                \centering
                \includegraphics[width=\textwidth]{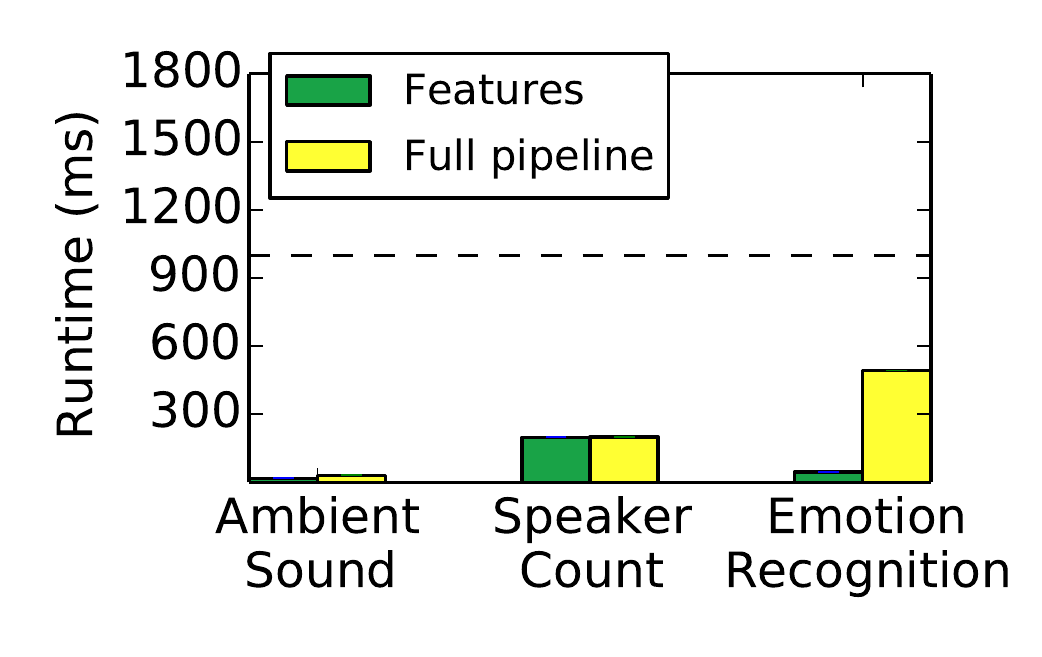}
                \small{(a) CPU}
        \end{minipage}%
        \hspace{0.01cm}
        \hfill
        \begin{minipage}[b]{0.23\textwidth}
                \centering
                \includegraphics[width=\textwidth]{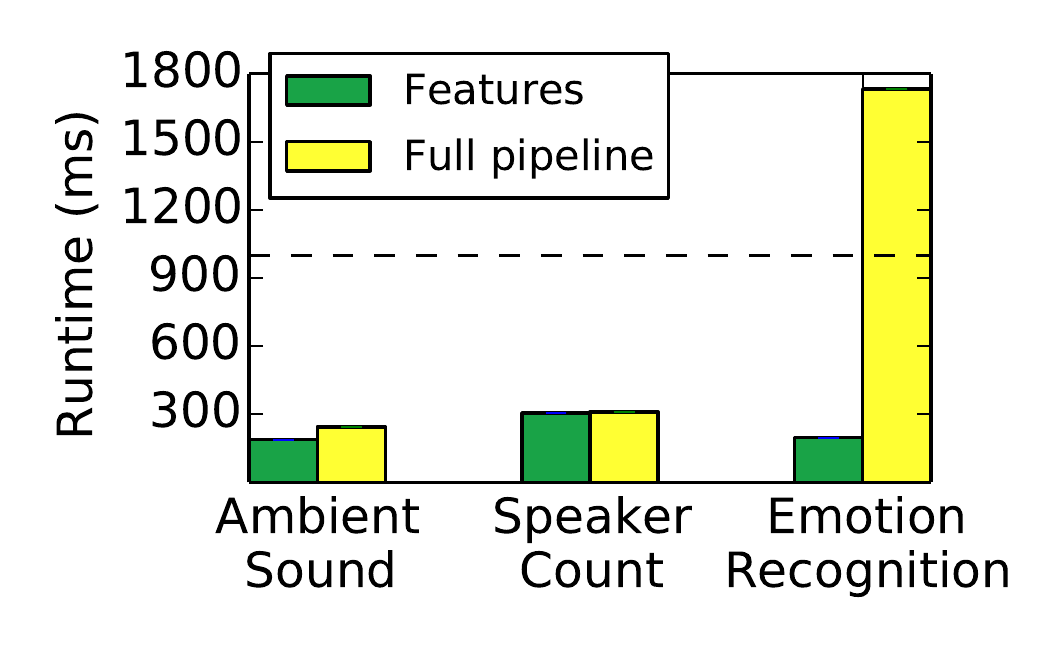}
                \small{(b) DSP}
        \end{minipage}
        \vspace{+0.2cm}
        \caption{Normalized runtimes per one second of audio sensing for the various applications when execution happens on (a) the CPU and (b) the co-processor.}
        \label{fig:runtime}
        \vspace{-0.5cm}
\end{figure}

The execution of any classification pipeline of the system is preceded by the admission filters ensuring that no unnecessary additional processing is incurred. Since they are always applied to the audio input, being capable of running fast and efficiently is of prime importance. In Table~\ref{table:runtime_filters} we demonstrate that although the DSP is between $5$ and $7$ times slower than the CPU, the admission filters occupy a small $0.045$-$0.066$ fraction of the audio processing per second. This spares execution space for the more compute-heavy scenarios the normalized runtime of which we display in Figure~\ref{fig:runtime}.

It is worth pointing out that since Figure~\ref{fig:runtime} portrays normalized runtimes per one second of audio sampling, we can easily distinguish between the application scenarios that can be run in real time on either of the two processing units (CPU and DSP). The quad-core Krait CPU performs all tasks in real time, while for the DSP the emotion recognition ($14$ GMMs) and speaker identification ($22$ GMMs) use cases are bottlenecks. The PLP feature extraction stage shared by the two application scenarios, however, consumes less than $12\%$ of the total emotion recognition execution making these features computation real-time on the DSP. The rest of the applications can easily run all of their stages on both the CPU and the DSP without introducing delays.


A noticeable difference between the two processing units is that the DSP operates between $1.5$ and $12$ times slower than the CPU on the various classification scenarios. The variability in the slow-down factor hints that the DSP treats the computation of the acoustic features differently from the CPU. This is because although the DSP supports floating point operations, some of them such as division and square root are implemented in software. In contrast, modern processors have dedicated hardware instructions for the same set of operations. The pitch computation algorithm, for instance, which is quadratic in the number of the samples in a frame, is composed predominantly of additions and multiplications which favor the co-processor hardware. As we can see from the figure, the DSP is only $1.5$ times slower than the CPU on the speaker count feature set where the pitch estimation dominates the MFCC computation. On the other hand, the ambient features computation on the DSP takes slightly over $11$ times longer than the CPU. The basic operations for these features include not only floating point division, but also taking the square root and performing discrete cosine transforms, all of which incur a processing overhead.


\subsubsection{Power Consumption}

\begin{table}[t]
\centering
\small
\begin{tabular}{rrr}
& \textbf{CPU} & \textbf{DSP} \\
\hline
\hline
\textbf{Silence} & 12.23 mW & 1.84 mW\\
\textbf{Speech} & 17.61 mW & 2.54 mW\\
\hline
\end{tabular}
\vspace{+0.2cm}
\caption{Normalized average power consumption in mW for the silence and speech admission filters.}
\label{table:power_filters}
\vspace{-0.4cm}
\end{table}

\begin{figure}[t]  
        \begin{minipage}[b]{0.23\textwidth}
                \centering
                \includegraphics[width=\textwidth]{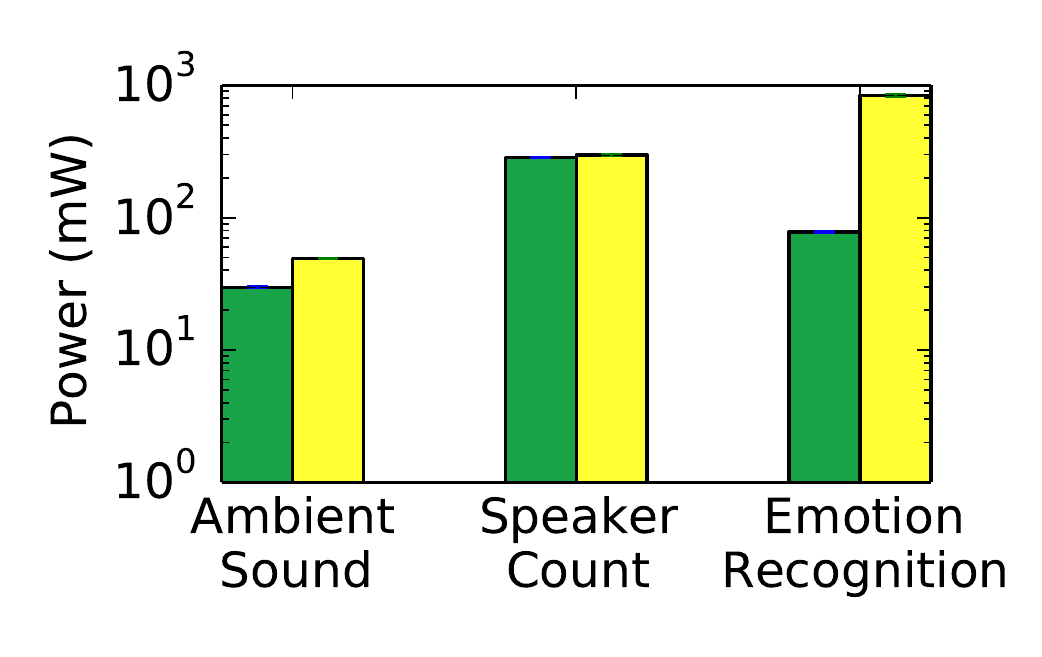}
                \small{(a) CPU}
        \end{minipage}%
        \hspace{0.01cm}
        \hfill
        \begin{minipage}[b]{0.23\textwidth}
                \centering
                \includegraphics[width=\textwidth]{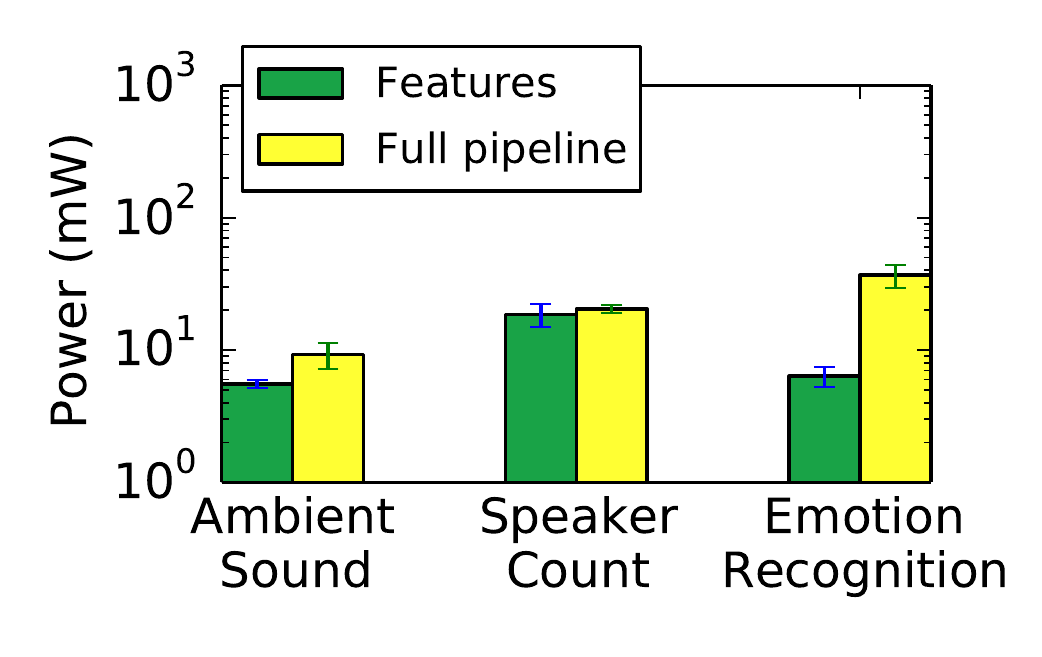}
                \small{(b) DSP}
        \end{minipage}
        \vspace{+0.2cm}
        \caption{Average power consumed by the various applications when running on (a) the CPU and (b) the co-processor. Values on the Y axis are in a logarithmic scale.}
        \label{fig:power}
        \vspace{-0.5cm}
\end{figure}

In this subsection, we provide critical insights on the relative efficiency with which the DSP is able to perform the various algorithmic tasks compared to the CPU. The measurements have been obtained through a Monsoon Power Monitor \cite{powerMonitor}. The reported values account only for the processing required by the algorithms without including the cost of maintaining the CPU awake and pulling audio data from the microphone sensor. This is done for this subsection only and so that we can better compare and contrast the energy overhead incurred by the pipeline stages themselves. The average power consumed by maintaining a wake lock on the CPU with a screen off on the MDP device is $295$mW, while keeping the microphone on adds around $47$mW on top for a total of $342$mW which is consistent with the reported values by Lu et al. \cite{Lu:2011:SEE:2021975.2021992}. The sampling of one microphone on the DSP with $8$kHz maintains a current of $0.6\sim0.9$mA ($2\sim4$mW) which is comparable to other sensors on low-power chips \cite{LittleRock}. Since continuously sampling the microphone on the DSP is not a publicly released functionality yet, we have obtained the exact values for the MDP board through the Qualcomm support team.


As already mentioned in the previous section, the admission filters are performed always as a first step in the processing which is why it is also important for them to be energy efficient. Table~\ref{table:power_filters} shows this is indeed the case. The DSP is around $7$ times more energy efficient for both tasks than the CPU. The overhead of continuously checking for whether the environment is silent is negligibly small on the DSP as the power does not exceed $2$mW. In the non-silent case, the additional energy cost of performing a decision tree classification on the type of sound is very low on the DSP, being merely $0.7$mW on top of the first admission filter.


The stages of the application algorithms, on the other hand, are more compute-heavy and consume much more power (Figure~\ref{fig:power}). Despite this, \textit{the DSP is an order of magnitude more energy efficient than the CPU}. The emotion recognition for instance consumes $836$mW on average on the CPU, while this number drops to merely $37$mW for the full pipeline on the co-processor. For the emotion/speaker detection task the DSP is thus more than $22$ times more power efficient than the main processor. Similarly, the full speaker count algorithm requires an average of $296$mW of power on the CPU, whereas on the co-processor the same execution consumes barely $21$mW. To put these numbers into perspective, if we add the required power for maintaining the audio sensing on the CPU, a standard $2300$mAh battery would last less than $14$ hours if it performs only speaker counting on the mobile phone and nothing else. If we assume that the CPU drains the battery with $30$mW of power in standby state \cite{nexus5}, and the DSP microphone sampling consumes $4$mW on average, then adding the $21$mW of the DSP for the same speaker counting task means that the battery would last for more than $154$ hours if it only counts speakers on the co-processor.





\subsection{Optimization Benchmarks}
\label{sec:optimization_benchmarks}

In this section we elaborate on several important aspects of the system design related to the introduced optimization techniques. We comment on the expected trade-offs between the accuracy and the savings in the processing. 

\subsubsection{Admission Filters: Neutral Emotions}


Here we discuss the implications of adding an admission filter that attempts to disambiguate between neutral and other emotions as a step that intercepts the full emotion recognition pipeline. We recall that the filter uses $2$ Gaussian Mixture Models, one representative of neutral speech, and another one absorbing the rest of the emotions. Table~\ref{table:neutral} demonstrates that such a model achieves an overall accuracy of $76.62\%$ and a false negative rate, i.e. neutral emotions predicted as non-neutral, of around $23$\%. While false negatives unnecessarily trigger the full pipeline of evaluating non-neutral narrow emotion models, in practice, the introduction of such an admission filter early into the pipeline is worthwhile even with this level of inaccuracy because of the following reasons. First, the figures outperform the overall accuracy of $71\%$ demonstrated by EmotionSense on the full set of emotions. Second, as discussed by Rachuri et al. \cite{Rachuri:2010:EMP:1864349.1864393} neutral speech occurs between $60$\% and $90$\% of the time making the short-circuiting possible for the majority of emotion recognition use cases even when false negative errors occur.


\begin{table}[t]
\centering
\small
\begin{tabular}{rrr}
& \textbf{Neutral} & \textbf{Non-Neutral} \\
\hline
\hline
\textbf{Neutral} & 77.26\% & 22.74\% \\
\textbf{Non-Neutral} & 24.20\% & 75.80\% \\
\hline
\end{tabular}
\vspace{+0.2cm}
\caption{Confusion matrix for neutral vs. emotional speech. Results are obtained via 10-fold cross validation.} 
\label{table:neutral}
\vspace{-0.5cm}
\end{table}

\begin{figure}[t]
		\centering
        \begin{minipage}[b]{0.33\textwidth}
                \centering
                \includegraphics[width=\textwidth]{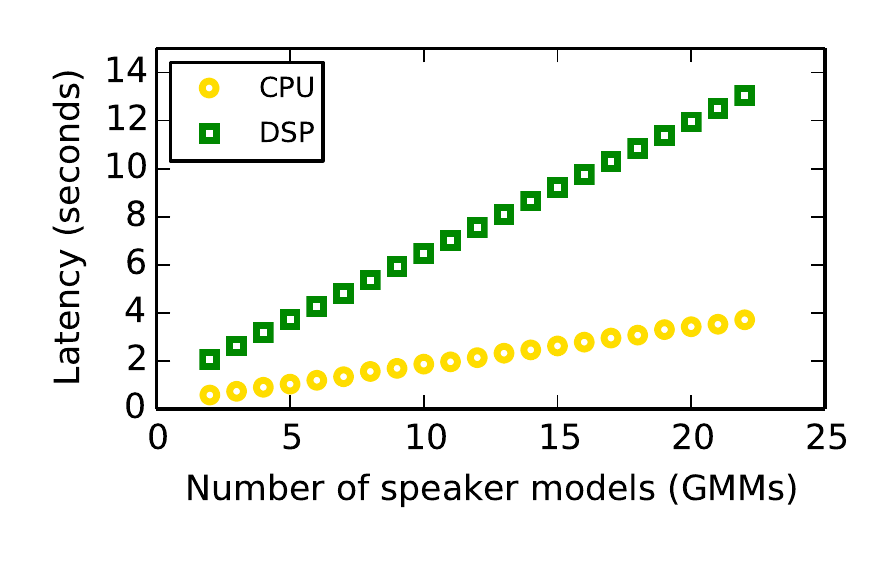}
        \end{minipage}%
        \caption{Runtime of the emotion and speaker recognition use cases as a function of the number of GMMs.} 
        \label{fig:gmm_runtime}
        \vspace{-0.4cm}
\end{figure}

The importance of the neutral emotion biasing becomes more pronounced when we take into account the following fact. Due to the DSP memory constraints which prohibit the deployment of more than $5$ emotion models on the DSP, the full pipeline needs to be executed on the power-hungry CPU. However, if the admission filter, which has $2$ GMMs, is deployed and executed fast on the DSP, for more than $60$\% of the time the emotion will be flagged as neutral and the processing will remain entirely on the low-power unit. In Figure~\ref{fig:gmm_runtime} we demonstrate that this scenario can be achieved. We plot the runtime of the speaker/emotion recognition pipeline as a function of the number of speaker/emotion models (GMMs) involved in the classification step. 
As can be seen from the figure, a pipeline with $2$ GMMs, which corresponds to the neutral emotion biasing, can be executed in less than $5$ seconds which is the EmotionSense base audio sampling period. In other words, this step can be executed in real time on the DSP and the emotion processing can indeed remain there for the majority of use cases given the dominance of neutral speech in everyday conversation settings.

Furthermore, when the emotion is flagged as non-neutral the further processing needs to occur only on the set of narrow models comprising the other broad emotions (happy, sad, afraid, angry). Thus, the revised full pipeline leads to the evaluation of the likelihood for a total of $10$ GMMs ($2$ for the admission filter plus $8$ for the narrow non-neutral emotions) as opposed to $14$ in the original version of EmotionSense.


\subsubsection{Locality of Sound Classification}
\begin{figure}[t] 
		\centering
        \begin{minipage}[b]{0.23\textwidth}
        		\centering
                \includegraphics[width=\textwidth]{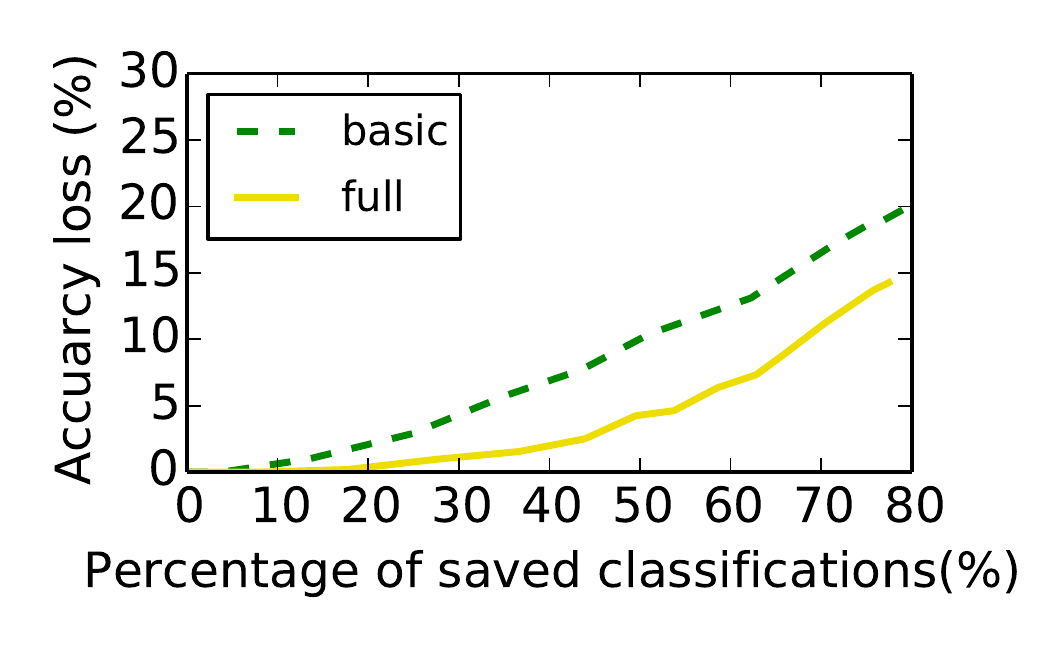}
		        \small{(a) Ambient sounds}
        \end{minipage}%
        \hfill
        \begin{minipage}[b]{0.23\textwidth}
        		\centering
                \includegraphics[width=\textwidth]{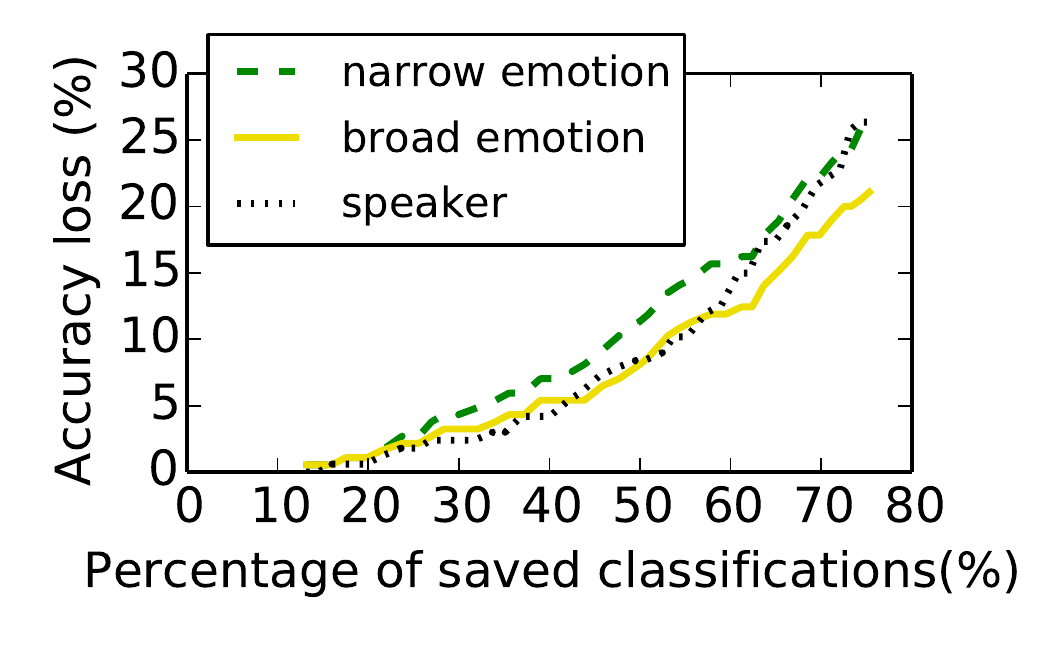}
                \small{(b) Emotions/Speakers}
        \end{minipage}%
        \vspace{+0.2cm}
        \caption{The percentage of misclassified sounds/ emotions/ speakers as a function of the proportion of saved GMM classifications due to engaging the similarity detectors. The similarity between subsequent sounds is computed based on two cases for the ambient sounds: the basic (without MFCC) and full set of features (with MFCC).}
        \label{fig:ambient_similarity}
        \vspace{-0.5cm}
\end{figure}

In this part of the analysis we shed light on the consequences of adding similarity detectors to the processing pipelines. We recall that we exploit behavioral locality so that when acoustic features from neighboring audio windows are sufficiently similar, classifications are bypassed and the sound category label from the previous inference is propagated to the next window. This optimization introduces false positive errors when features from subsequent windows are similar but the sound categories are not the same. 

In Figure~\ref{fig:ambient_similarity}(a) we plot the proportion of similarity false positives as a function of the saved GMM classifications when the similarity detector component is introduced into the ambient processing pipeline. To obtain the figure we vary the similarity distance threshold between the vectors representing subsequent acoustic fingerprints. The threshold in the figure is implicitly represented by the number of saved computations where higher thresholds result in more saved classifications and larger errors. We notice that when we exploit the full feature set with MFCCs (Table~\ref{table:features}) to compute the similarity between subsequent audio windows we can save around $50\%$ of the classifications at the expense of a $4.3\%$ penalty in the classification accuracy. 


Figure~\ref{fig:ambient_similarity}(b) demonstrates the accuracy loss incurred by wrongly marking subsequent audio windows with the same broad emotion, narrow emotion and speaker label when the similarity distance between speech features falls below a threshold and classifications are omitted. A noticeable observation is that the accuracy penalty is higher than what has been achieved by the ambient sound similarity component. This can be explained by the fact that the inherent difference in acoustic fingerprints is greater for ambient sounds than expressed emotions. In general, subtle changes in valence are difficult to detect. Nevertheless, on the broad emotion recognition task we are still able to save 20\% of the classifications at the expense of an accuracy loss of only $1\%$. 


\subsubsection{Selective CPU Offloading}

\begin{figure}[t] 
		\centering
        \begin{minipage}[b]{0.33\textwidth}
                \includegraphics[width=\textwidth]{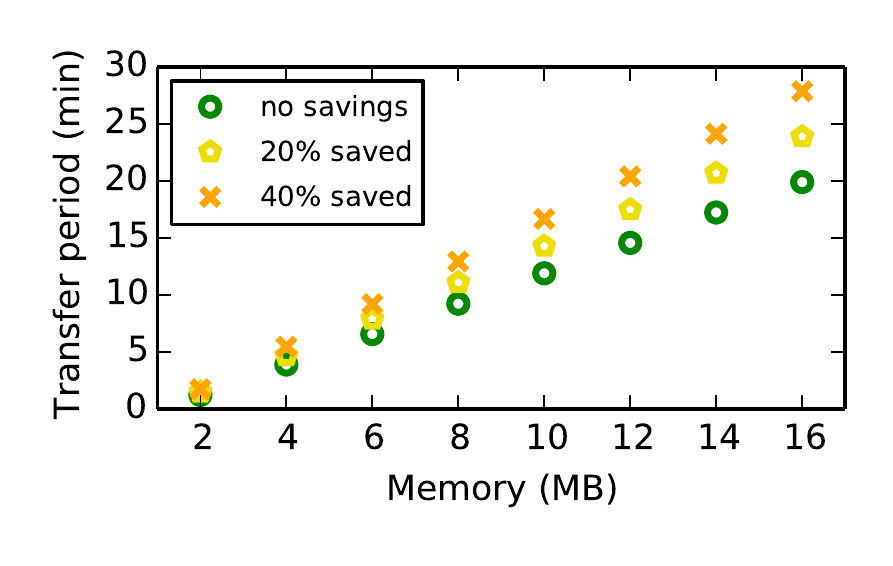}
        \end{minipage}%
        \caption{Time in minutes before the CPU is woken up by the DSP as a function of the runtime memory constraint on the co-processor. The additional gain in time is also given when the similarity detectors eliminate a percentage of the classifications by propagating class labels, discarding features and thus freeing space.} 
        \label{fig:cpu_wakeup_frequency_memory}
        \vspace{-0.5cm}
\end{figure}

As discussed in Section~\ref{sec:offloading} the DSP is subject to runtime memory constraints which affects the rate at which the CPU is approached to perform further processing on accumulated acoustic features. The main sources of space concerns are the PLP features occupying nearly $64$KB when extracted once every 5 seconds and any ambient features remaining unlabeled because of the unavailability of the full set of sound models on the DSP. As already discussed the intensity of detected speech is the dominating factor in how often the CPU is woken up by the DSP. To provide a lower bound on the amount of time spent before the CPU needs to be interrupted from its sleep mode, we consider the case when human voice is continuously detected.

In Figure~\ref{fig:cpu_wakeup_frequency_memory} we plot the time the DSP can be actively processing speech audio data before running out of space and waking up the CPU to transfer acoustic feature buffers. In the current mobile development platform the memory limit is $8$MB which results in the CPU being approached for processing by the DSP once every $9$ minutes assuming no emotion/speaker classification savings are allowed and no accuracy losses are incurred. Involving the similarity detectors into the pipeline leads to propagating a proportion of the class labels and discarding the corresponding features which frees up space. We can thus extend the wake-up time to $13$ minutes at the expense of a modest $5$\% loss in accuracy when $40$\% of the features are discarded because of class label propagation. 
Note that delaying the transfers of the feature buffers to the CPU is desirable energy-wise since the average power consumed to wake it up is generally high. We measure $383$mW on average on the MDP during the wake-up process which may last several seconds. In addition, the CPU does not go immediately to low-power standby mode and typically on production phones the CPU may remain idle for $10$-$20$ seconds (such as Samsung Galaxy S, S2, S4) after the processing is over which incurs a power consumption overhead.



\subsection{Full System Evaluation}
\label{sec:full_eval}

In this subsection we provide an exhaustive evaluation of the full system given various workload settings and latency requirements. 

\subsubsection{Assumptions}
Based on the analysis performed in the previous sections, we assume default parameters for several of the system components. We activate the similarity detectors so that the percentages of saved classifications are $50\%$, $40\%$ and $20\%$ for ambient sounds, speakers and emotions respectively. This is done so that we maintain a reasonable accuracy loss of $4\%-5\%$ for the ambient sounds and speakers and $1\%$ for the emotions. Given the detailed experiments on the distribution of emotions performed by Rachuri et al. \cite{Rachuri:2010:EMP:1864349.1864393}, we expect neutral emotions to be encountered around $60$\% of the time which is when the DSP is capable of performing the entire emotion processing. We use the mobile system parameters of a popular smartphone, Google Nexus 5, featuring the Qualcomm Snapdragon $800$ platform and having a battery of capacity $2300$mAh. The estimated standby time is officially reported to be up to $300$ hours which translates to an average standby power of $30$mW \cite{nexus5}. Based on measurements we performed on the MDP and other popular smartphones such as Samsung Galaxy S, S2 and S4 we assume the CPU remains idle for $15$ seconds after all processing is over and before going to deep sleep (standby) mode. By default, we wake up the CPU once every $11.1$ minutes when speech is detected as we run out of space given a runtime memory constraint for the DSP of $8$MB. Last but not least, the cross-pipeline optimization of tagging the speech features with the detected gender allows us to reduce the number of speaker models against which we evaluate the likelihood of the speech features. Our assumption is that the gender detection is able to eliminate roughly half of the speaker models which leaves us with $11$ GMMs given our test dataset of $22$ speakers.

\subsubsection{Unconstrained Battery Usage}

Here we give an overview of how the system drains power given that the full battery capacity is available for use solely by the system. The goal is to contrast how the system fares against baselines, whereas estimates for realistic battery drains under common workloads are given in the next section. We compare the model against three baselines including a CPU-only solution, and two solutions for the CPU and DSP respectively without the introduced improvements of similarity detectors, neutral emotions admission filter and cross-pipeline optimizations. We vary the distribution of sound types a user encounters in everyday settings to demonstrate the level of dependence of the system on the type of processing performed (voice/ambient cases). We fix the proportion of silence in a day to $1/3$ of all types of sounds which corresponds to roughly $8$ hours of a night's sleep. 

\begin{figure}[t] 
		\centering
        \begin{minipage}[b]{0.33\textwidth}
                \includegraphics[width=\textwidth]{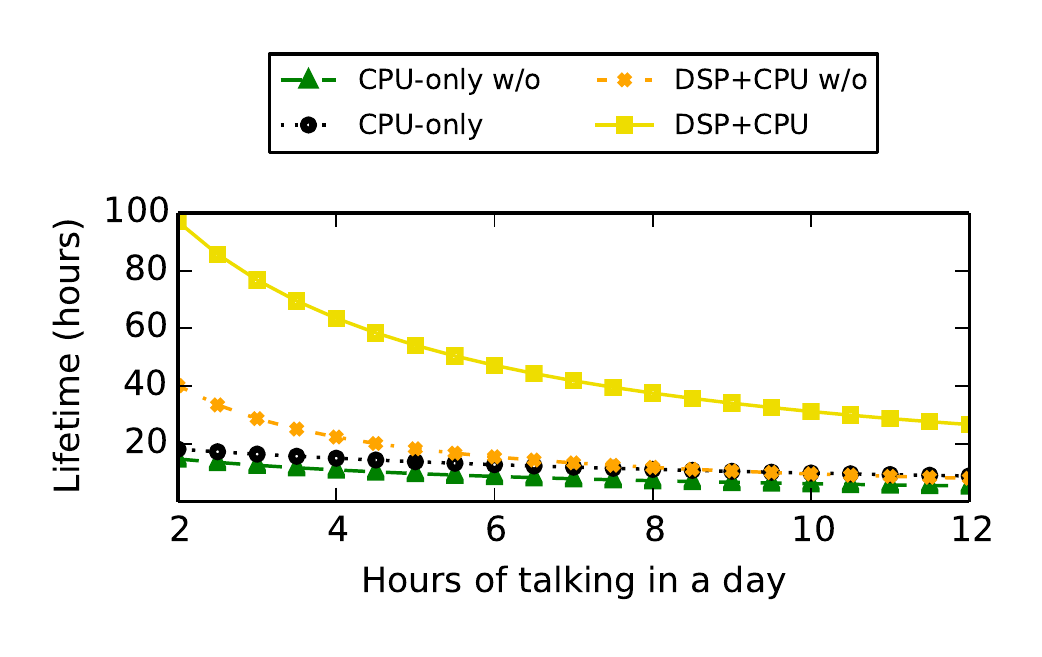}
        \end{minipage}%
        \caption{Lifetime in hours of the system running on the CPU only or the DSP+CPU as a function of the proportion of detected speech during a $24$-hour day. Two versions of the system are provided: without (w/o) and with optimizations.} 
        \label{fig:baselines_workload}
        \vspace{-0.3cm}
\end{figure}

In Figure~\ref{fig:baselines_workload} we vary the amount of detected speech during the day as the voice processing has the most pronounced effect on the battery given that the execution there relies on the heavier pipelines of the system. A first observation is that the DSP solution with optimizations is between $3$ and $7$ times more power-efficient than the CPU-only solutions. With unconstrained battery usage and $4.5$ hours of talking per day the integrated system with optimizations is able to last almost $60$ hours exceeding considerably the $14.4$ hours reached by a CPU-only solution with the same optimizations. Note that the targeted $4.5$ value is an average number of hours spent in conversations in a day as found by Lee et al. \cite{Lee:2013:SEF:2462456.2465426}. As we increase the amount of detected speech from $4.5$ to $12$ hours per day, the system longevity significantly decreases where we witness a $58$\% drop in the total hours of runtime for the DSP case and $41\%$ drop for the CPU-only case. 

Another insight is that the optimizations that have been introduced provide a noticeable improvement in the battery lifetime, especially for the DSP case where for $4.5$ hours of talking the total hours jump from $20$ to $60$. The optimizations are so crucial that we observe the following phenomenon: using the DSP without them reaches a point at around $8$ hours of speech where the CPU + co-processor design is less efficient than simply having the optimizations on the CPU. This is expected since the major energy burden are the emotion recognition and speaker identification classifications which always run on the CPU. In this case, running the optimization procedures on the DSP is critical for enabling the truly continuous sensing of the microphone. The battery is able to last $2$ to $3$ times more if the mentioned optimizations are added to the DSP+CPU solution.

\subsubsection{CPU Workload Analysis}

\begin{table}[t]
\centering
\small
\begin{tabular}{|l|l|}
\hline
\textbf{Category} & \textbf{Examples of Profiled Applications} \\
\hline
\hline
 Books & Amazon Kindle, Bible \\
 \hline
 Browsing \& Email & Firefox, Yahoo! Mail \\
 \hline
 Camera & Camera, Panorama 360 Camera \\
 \hline
 Games & Angry Birds, Fruit Ninja \\
 \hline
 Maps \& Navigation & Google Maps, Street View \\
 \hline
 Media \& Video & Android Video Player, VPlayer \\
 \hline
 Messaging & GO SMS Pro, KakaoTalk \\
 \hline
 Music \& Audio & n7player Music Player, Winamp \\
 \hline
 Photography & Adobe Photoshop Express, Photo Editor \\
 \hline
 Social & Facebook, Twitter \\
 \hline
 Tools \& Productivity & Advanced Task Killer, Easy Battery Saver \\
 \hline
 Other & Skype, Super Ruler Free \\
\hline
\end{tabular}
\vspace{+0.2cm}
\caption{Categories of smartphone apps used in the CPU workload evaluation.} 
\label{table:app_category}
\vspace{-0.5cm}
\end{table}

In this final subsection we study the implications of running DSP.Ear together with other common workloads generated by smartphone users. For this purpose, we use a dataset provided by the authors of AppJoy \cite{AppJoy}. It consists of traces of hourly application usage from $1320$ Android users, and is collected between February---September $2011$ as part of a public release on the Android marketplace. The number of applications found in the user traces exceeds $11$K which renders the energy and CPU workload profiling of all apps impractical. Therefore, we group the apps with similar workload characteristics into categories, as shown in Table~\ref{table:app_category}, and profile several typical examples from each category. To obtain measurements we run $1$-$2$ minute interactive sessions (open-run-close) with each app using the Trepn Profiler \cite{Trepn} for the CPU load and the Power Monitor \cite{powerMonitor} for the power consumption.

\begin{figure}[t] 
		\centering
        \begin{minipage}[b]{0.23\textwidth}
        		\centering
                \includegraphics[width=\textwidth]{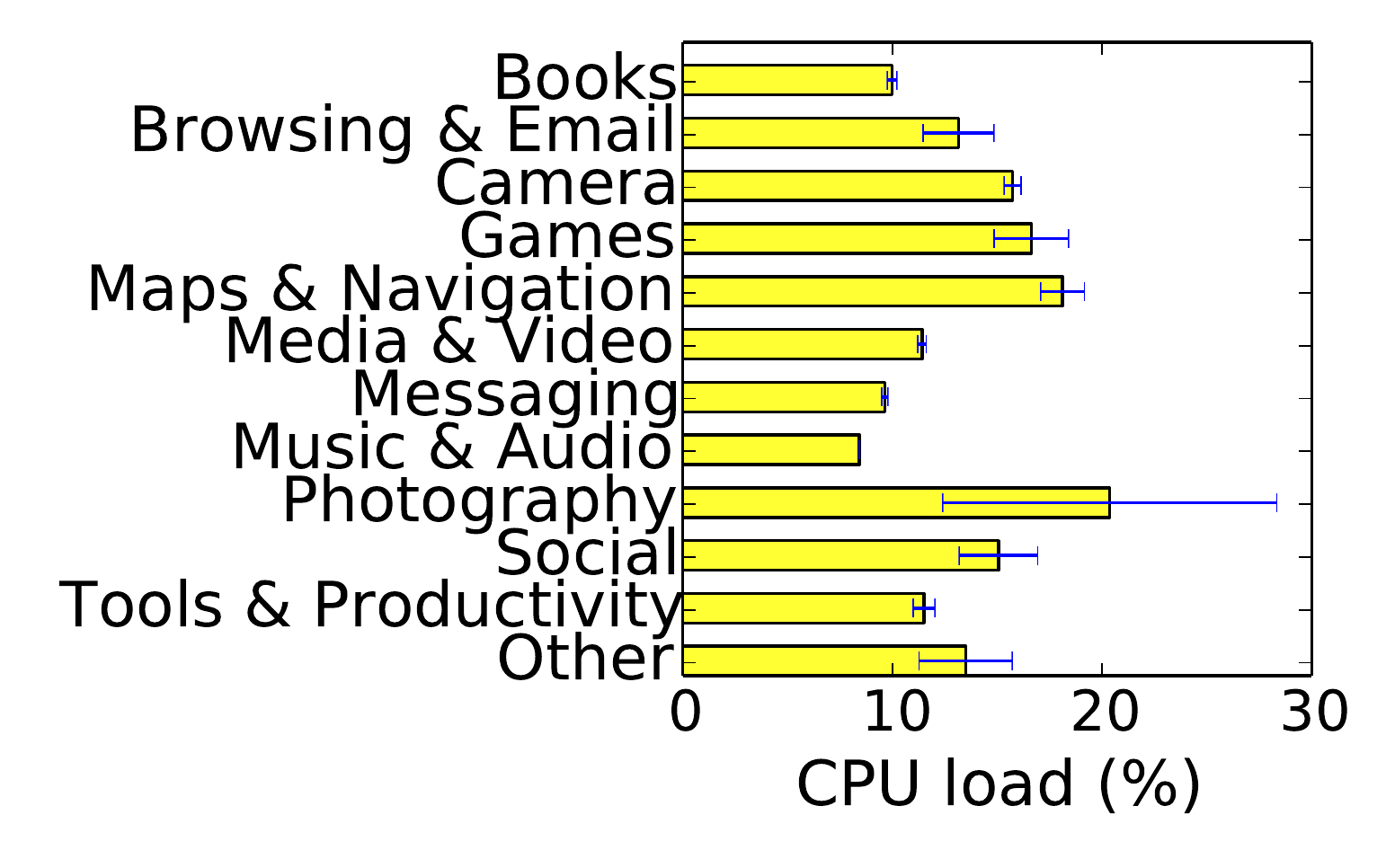}
		        \small{(a) CPU load (\%)}
        \end{minipage}%
        \hfill
        \begin{minipage}[b]{0.23\textwidth}
        		\centering
                \includegraphics[width=\textwidth]{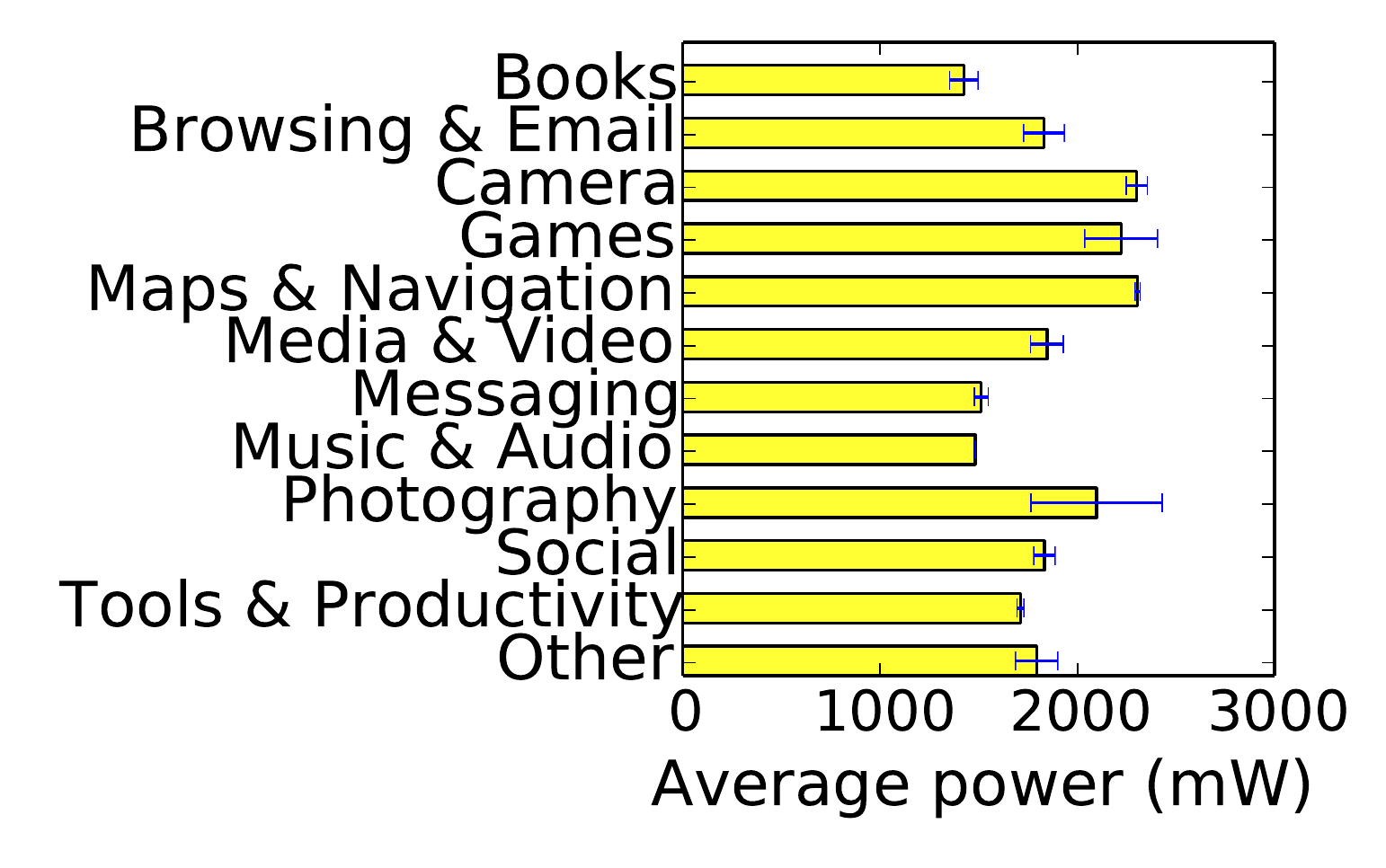}
                \small{(b) Power (mW)}
        \end{minipage}%
        \vspace{+0.2cm}
        \caption{The average (a) CPU load and (b) power of interactively (screen on) running applications from the given categories.} 
        \label{fig:categories}
        \vspace{-0.5cm}
\end{figure}

There are two major issues to consider when analyzing how the continuous audio sensing workloads incurred by our system interact with the smartphone usage patterns. The first consideration is the overhead of waking up the CPU from the DSP to offload emotion and speaker recognition. When $4.5$ hours of conversations are encountered on average in a day \cite{Lee:2013:SEF:2462456.2465426}, the  overhead of the wake-up operations plus keeping the CPU in active state \textit{excluding} the sound classification processing itself is $4.5\%$ of the battery capacity. This amounts to $12$ minutes of sacrificed web browsing energy-wise. In practice, the CPU will be already active for other tasks at times. Exactly when our processing will take advantage of this is a complex function not only of the CPU active state encounters, but also of the unknown distribution of speech encounters. On average the users in our dataset spend $72$ minutes in interactive sessions, whereas the total emotion and speaker processing for $4.5$ hours of conversations lasts $82$ minutes. Since the $4.5$\% wake-up overhead is not critical for our further analysis, we make a simplifying assumption that half of the time the CPU will be on for other tasks when the speech pipelines engage it.



The second primary consideration is the energy and CPU overhead of running the speech processing on the CPU together with other active apps on the phone. A major burden is the interactive usage when the screen is on \cite{Carroll:2010:APC:1855840.1855861}, where the display energy overhead is attributed to the LCD panel, touchscreen, graphics accelerator, and backlight. The Android OS  maintains one foreground app activity at a time \cite{AppJoy}, making the application currently facing the user and the screen the main sources of energy consumption. When running the audio pipelines on the Snapdragon MDP device, we observe that the power consumption is additive as long as the normalized CPU load (across cores) remains below $80$\%. The apps from the various categories as shown in Figure~\ref{fig:categories}(a) rarely push the CPU beyond $25$\% in interactive mode. 
The additional CPU load of performing the speaker/emotion identification is $15$\% for a total of $40$\% cumulative utilization. This remains below the threshold beyond which the energy consumption stops being additive and outgrows the value under normal workloads. 

\begin{figure}[t] 
		\centering
        \begin{minipage}[b]{0.33\textwidth}
                \includegraphics[width=\textwidth]{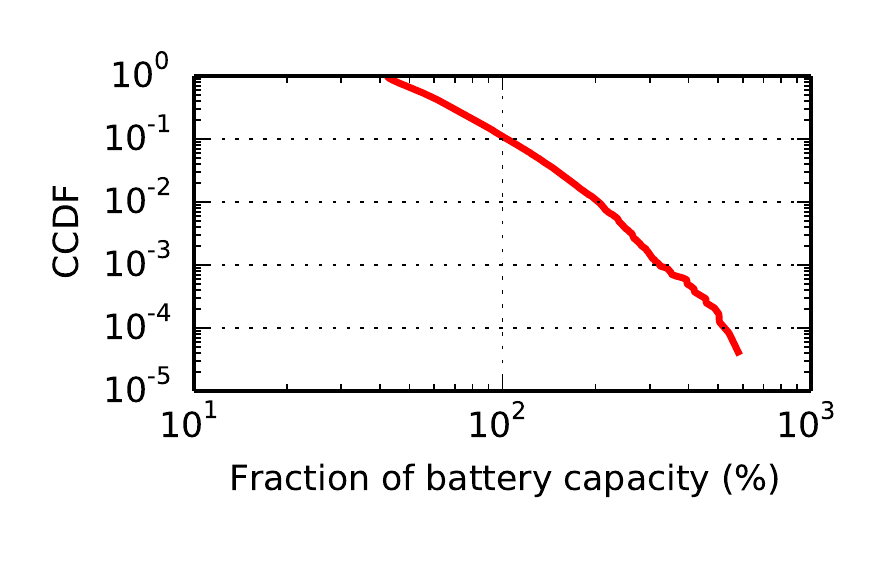}
        \end{minipage}%
        \vspace{-0.2cm}
        \caption{Complementary cumulative distribution function (CCDF) of the percentage of battery capacity drained in $24$ hours of smartphone usage.} 
        \label{fig:ccdf_usage}
        \vspace{-0.5cm}
\end{figure}

To evaluate the overhead of running our system together with other smartphone usage workloads, we replay the application traces from the AppJoy dataset with measurements performed on the Snapdragon MDP device. In Figure~\ref{fig:ccdf_usage} we plot the percentage of the battery capacity ($2300$mAh) expended in $24$ hours of operation. Since the dataset does not provide explicit information for background application usage when the screen is off, the figure accounts for the interactive usage of the phone, our system running in the background and the CPU standby power draw. This is a best case scenario for our system where third-party background services are disabled, and periodic server data synchronization and push notifications are off. \textit{In this scenario, in $90$\% of the days 
the battery is able to sustain at least a full day of smartphone usage without charging}. This is encouraging news, since for the significant proportion of daily usage instances, users need not worry about their everyday charging habits. 

The next scenario we focus on is adding the overhead of running typical background services such as the ones started from social networking applications, mail clients, news feeds, and music playback. In the AppJoy dataset $50$\% of the users have installed Facebook on their phone, $69$\% have Gmail or Yahoo! Mail, $10$\% have Twitter and $15$\% have Pandora radio or Winamp. Social networking and microblogging apps such as Facebook and Twitter sync user timelines periodically in the background with default update periods of $1$ hour or $15$ minutes respectively. Mail clients, on the other hand, usually adopt push notifications as their primary method of synchronization where cached data on the phone is updated through the Google Cloud Messaging services only when new e-mails are received on the mail server. Since these services trigger operations either periodically or based on some external events, the CPU may not be awake when the updates occur.

\begin{figure}
  \begin{minipage}[b]{0.56\linewidth}
    \centering
    \includegraphics[width=\linewidth]{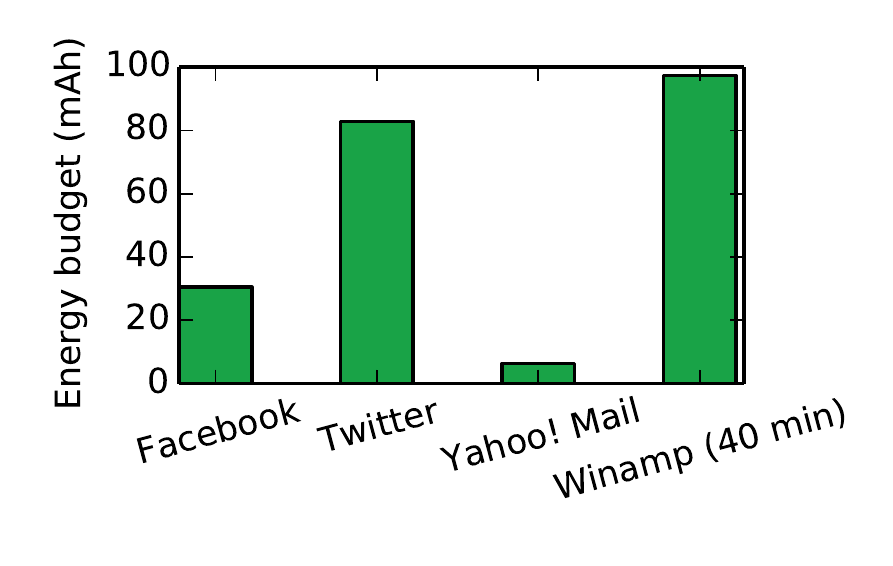}
    \par\vspace{0.0pt}
  \end{minipage}%
  \begin{minipage}[b]{0.40\linewidth}
  	\scriptsize
    \centering

	\begin{tabular}[b]{ |l|l| }
	  \hline
	  \textbf{Apps} & \textbf{Background Load} \\
	  \hline \hline
	  Facebook & $232$KB per $1$ hour \\ \hline
	  Twitter & $54.5$KB per $15$ min \\ \hline
	  Yahoo! Mail & $513$ KB, $7$ e-mails \\ \hline
	  Winamp & $40$ min of playback \\ \hline
	\end{tabular}
	\par\vspace{4.5em}
\end{minipage}
\caption{Energy budget expended in a day by the background services of $4$ popular mobile apps. Measurements are obtained with the Snapdragon MDP where server data from Facebook, Twitter and Yahoo! Mail is downloaded via WiFi.}
\label{fig:services}
\vspace{-0.5cm}
\end{figure}

In Figure~\ref{fig:services} we present the energy budget required by $4$ representative applications/services so that they are able to run in the background over a day with default sync settings. The measurements account for the worst case when the CPU needs to be woken up on every occasion any of the services needs an update. Given this and an extrapolated $40$ minutes of background music playback per day, the total energy budget remains below $10$\%. \textit{If we aggressively add this workload to all daily traces, we find that in more than $84$\% of the instances, the users will not need to charge their phone before the full day expires.} 
This is again encouraging, since even when users have several popular apps running in the background, they can, in a considerable proportion of the cases, engage in uninterrupted audio life logging.

\section{Limitations and Future Work}
\label{sec:discussion}


In what follows, we outline key limitations of our implementation and sketch potential directions for future work.

\vspace{+0.5em} \noindent \textbf{Programmability.} Due to imposed limitations on the supported devices by the publicly released APIs, the development of the system cannot be performed directly on a standard commodity phone with the same Snapdragon 800 processor. While this work is a proof-of-concept implementation through which we provide generalizable techniques, questions of programmability are still prominent. 

\vspace{+0.5em} \noindent \textbf{DSP Efficiency.} The current implementation does not fully take advantage of the more advanced DSP capabilities such as optimized assembly instructions or fixed point arithmetic. The primary advantage of using the C programming language with floating point operations for the DSP system implementation is that it allows for the prototyping of more advanced algorithms and importing legacy code. 
However, the challenges in supporting advanced algorithms are not solely based on the need to perform floating point operations or to reduce runtime through assembly optimizations. Limitations still remain --- for example, encoding the range of necessary models for each type of supported classification places significant strain on memory reserved for program space (e.g., speaker identification requires one separate model for each speaker recognized).

\vspace{+0.5em} \noindent \textbf{Scheduling.} Through our work we have explored a scheduling scheme with relatively short delays in selectively offloading tasks to the CPU. While opportunistically performing computation on the DSP at times when it is not busy might be generally more energy friendly than waking up the CPU, the issues of co-processor memory constraints persist. Further, the near real-time nature of various application scenarios requiring immediate feedback increases the need for short delays in the processing. Examples are behavior interventions such as distress detection, or detecting dangerous situations such as falling down the stairs. 

\vspace{+0.5em} \noindent \textbf{Future Work.} Potential directions for future work include cloud offloading, GPU offloading and substituting classification algorithms to trade-off accuracy vs. latency. Performing the expensive pipeline classification tasks in the cloud has the potential to substantially improve the energy profile of sensor inference systems on mobile phones as long as the right balance is found between the maximum tolerated latency and the overhead of uploading features. Cloud offloading with short delays will have a high energy overhead, therefore, accumulating features for larger batch transfers will be highly desirable. Similar concerns for batched execution with larger delays are valid for the GPU as well where the overhead in approaching the GPU stems from the CPU-GPU memory transfer costs. A suitable workload for the GPU would be SIMD-like processing where one and the same task is performed in parallel on multiple data items, such as speaker recognition on multiple audio windows. 

\section{Design Observations}
In this section, we summarize the broader lessons learned from developing our system and draw guidelines for developers of audio sensing applications. There are several major design options to be taken into consideration when providing sensing solutions for DSPs.

\vspace{+0.5em} \noindent \textbf{Maximize pipeline exposure to the DSP}. The first primary design concern is related to the introduction of optimization techniques that allow the pipeline processing and inferences to occur predominantly on the low-power unit with minimal assistance from the CPU. This is important since the energy consumption on the DSP is usually orders of magnitude lower than the CPU. Example techniques we have leveraged to achieve this design goal are: 1) admission filters that are efficiently executed in the early pipeline stages on the DSP to interrupt unnecessary further processing of sensor data; and 2) behavioral locality detectors that propagate class labels when the acoustic fingerprints of nearby audio windows are sufficiently similar.

\vspace{+0.5em} \noindent \textbf{Favor decomposable classification algorithms}. The classification/inference stage of the pipelines is often the most computationally demanding algorithmic component. As we have seen with the emotion and speaker identification scenarios, it is a processing bottleneck for the DSP not only because of its extended latency but also because of the memory constraints prohibiting the deployment of models with a large number of parameters. If the classification stage could be \emph{decomposed into a series of incremental inferences}, it would allow part of the inferences to be computed efficiently on the DSP. A prominent example is the emotions recognition use case where the classification consists of matching features against each of a set of emotional models. We are able to exploit this property to quickly make a binary decision on the DSP whether or not the current emotion is the commonly encountered neutral speech. Only if the inference suggests the emotion is non-neutral will other emotional models be later evaluated and the higher-power CPU approached.

\vspace{+0.5em} \noindent \textbf{Minimize frequency of DSP-CPU interactions}. When the CPU is frequently approached by the DSP for assistance, chances are the CPU will not be already awake on all occasions and will need to be woken up from its deep sleep mode. As demonstrated earlier, this is accompanied with an energy overhead originating from both the wake-up operation itself and the trailing energy consumed before the CPU goes back to its low-power mode. We have demonstrated how to take advantage of the behavioral locality detectors to reduce the number of DSP-CPU contact points by prolonging the time before the next DSP buffer transfer.

\vspace{+0.5em} \noindent \textbf{Favor compact classification models}. Memory limitations of the DSP are prominent. The less space the classification models occupy, the more memory becomes available at runtime for application inferences and extracted features. In addition, compact model representations with fewer parameters allow for a larger set of models to be deployed on the DSP. The cumulative benefit is the DSP becomes less dependent on the CPU for processing which increases energy efficiency.

\section{Related Work}
\label{sec:related}
Modern mobile phones embedded with a variety of sensors provide an opportunity to build applications that can continuously capture information about the phone user. 
Continuous sensing, however, drains the phone battery. To overcome the battery challenges, many solutions have been devised, which can be broadly classified into two categories: software-based and hardware-based approaches. 

Prominent software-based approaches include adaptive duty cycling~\cite{conf/infocom/ConstandacheGSCC09, jigsaw2010, eemss2009}, triggered sensing~\cite{Zhang:2012:AGO:2426656.2426674,pcs,nericell2008}, and exploiting relations between various contexts~\cite{ace2012}. Adaptive duty cycling techniques enhance the static duty cycling by adjusting the interval according to various metrics such as user's context, energy budget, or mobility. While these techniques enhance duty cycling efficiency, they do not completely fill the gap to perform always-on sensing. In order for these schemes to perform always-on sensing, especially when a continuous stream of on-going events need to be captured, they would need to keep the main CPU awake, which would reduce the battery life considerably. Using a low-power sensor to trigger a high-power one~\cite{nericell2008} and techniques that exploit the relation between context items~\cite{ace2012} are applicable to specific scenarios. An example is accelerometer-based triggering of GPS, but it might be difficult to generalize these schemes across different sensors. 

Existing work on hardware-based solutions include exploring co-processor computation~\cite{Lu:2011:SEE:2021975.2021992} or using phone peripheral devices~\cite{Verma:2012:ATM:2426656.2426677}. Lu et al. \cite{Lu:2011:SEE:2021975.2021992} present a speaker identification system that achieves energy efficiency by distributing the computation between a low-power processor and the phone's main processor. Verma, Robinson and Dutta~\cite{Verma:2012:ATM:2426656.2426677} present AudioDAQ, a mobile phone peripheral device that can be plugged to the headset port of a phone and that provides continuous capture of analogue sensor signals. Most of these solutions, however, are based on custom-build hardware whereas our work focuses on hardware that is already present in commodity smartphones. Moreover, the optimizations that we presented extend the battery life further by many folds, and are not present in most existing systems.

CloneCloud~\cite{cloneCloud} achieves energy efficiency and reduces latency by offloading a part of the execution from a local virtual machine on the phone to device clones running in the cloud. Rachuri et al.~\cite{sociableSense2011} build a system that distributes computation between the phone and the cloud to balance energy-latency trade-offs.  Although there is a similarity between these works and balancing computation between the main CPU and the co-processor, the challenges faced by these schemes are very different. While network related issues are an important consideration in the former, memory, computation, and energy limitations are the main focus in the latter.

\section{Conclusion}
\label{sec:conclusion}
In this paper we have studied the trade-offs of using a low-power co-processor found in state-of-the-art smartphones to perform continuous microphone sensing and logging of a variety of sound-related behaviors and contexts. 
We have developed DSP.Ear, an integrated system with multiple interleaved audio inference pipelines that are able to run continuously together $3$ to $7$ times longer than when deployed entirely on the CPU. 
Our system is also $2$ to $3$ times more power efficient than a na\"{i}ve DSP-based design. 
Further, we have introduced techniques such as \emph{admission filters}, \emph{locality of sound detectors}, \emph{cross-pipeline optimizations} and \emph{selective CPU offloading} that prove critical for lowering the power profile of near-real time applications such as behavior interventions. The insights drawn and the techniques developed in the presented work can help towards the growth of the next-generation of context-aware and reactive mobile applications, able to operate uninterrupted and concurrently without noticeably affecting the device battery lifetime.

\section{Acknowledgments}
This work was supported by Microsoft Research through its PhD Scholarship Program. We are grateful to the Qualcomm team for their cooperation. We would also like to thank Natasa Milic-Frayling for insightful discussions, as well as the anonymous reviewers and our shepherd, Rajesh Gupta, for helping us improve the paper.

\balance

\end{document}